%% file: main.tex
\documentclass[letterpaper,twocolumn,10pt]{article}
\usepackage{usenix-2020-09}
\def\BibTeX{{\rm B\kern-.05em{\sc i\kern-.025em b}\kern-.08emT\kern-.1667em\lower.7ex\hbox{E}\kern-.125emX}}

\usepackage{hyperref}

\usepackage[nohyperlinks, nolist]{acronym}
\usepackage{xspace}
\usepackage{soul}
\usepackage{pgf-umlsd}
\usepackage{algorithm}
\usepackage[noend]{algpseudocode}
\usepackage{tabularx}
\usepackage{calc}
\usepackage{dcolumn}
\usepackage{paralist}
\usepackage{enumitem}
\usepackage{cleveref}
\usepackage{amssymb}

\newcommand*\circled[1]{\tikz[baseline=(char.base)]{
		\node[shape=circle,draw,inner sep=1pt, style={text=white,fill=black}, font=\bfseries] (char) {#1};}}
\newcolumntype{d}[1]{D{.}{.}{#1}}

\DeclareMathAlphabet{\mathpzc}{T1}{pzc}{m}{it}
\newcommand{\sysname}{V'CER\xspace}
\newcommand{\ca}{$\mathpzc{CA}$\xspace}
\newcommand{\node}{$\mathpzc{Node}$\xspace}
\newcommand{\nodes}{$\mathpzc{Nodes}$\xspace}
\newcommand{\adv}{$\mathpzc{Adv}$\xspace}
\newcommand{\forest}{$\mathpzc{VF}$\xspace}
\newcommand{\primename}{Aggregator\xspace}
\newcommand{\primem}{$\mathpzc{Aggr}$\xspace}
\newcommand{\primeroot}{$\mathpzc{A_r}$\xspace}
\newcommand{\primets}{$\mathpzc{A_t}$\xspace}
\newcommand{\primeparity}{$\mathpzc{A_c}$\xspace}
\newcommand{\primesig}{$\mathpzc{A_s}$\xspace}

\newcommand{\epochtree}{$\mathpzc{E_T}$\xspace}
\newcommand{\epochroot}{$\mathpzc{E_r}$\xspace}
\newcommand{\epochs}{$\mathpzc{e}$\xspace}

\begin{acronym}
	\acro{ca}[CA]{Certificate Authority}
	\acro{crl}[CRL]{Certificate Revocation List}
	\acro{iot}[IoT]{Internet of Things}
	\acro{ocsp}[OCSP]{Online Certificate Status Protocol}
	\acro{pki}[PKI]{Public Key Infrastructure}
	\acro{poi}[PoI]{Proof of Inclusion}
	\acroplural{poi}[PoIs]{Proofs of Inclusion}
	\acro{mht}[MHT]{Merkle Hash Tree}
	\acro{smt}[SMT]{Sparse Merkle Tree}
	\acro{ct}[CT]{Certificate Transparency}
\end{acronym}

\begin{document}

\date{}
\title{V'CER: Efficient Certificate Validation in Constrained Networks}

\author{
	{\rm David Koisser}\\
	Technical University Darmstadt\\
	david.koisser@trust.tu-darmstadt.de
	\and
	{\rm Patrick Jauernig}\\
	Technical University Darmstadt\\
	patrick.jauernig@trust.tu-darmstadt.de
	\and
	{\rm Gene Tsudik}\\
	University of California, Irvine\\
	gene.tsudik@uci.edu
	\and
	{\rm Ahmad-Reza Sadeghi}\\
	Technical University Darmstadt\\
	ahmad.sadeghi@trust.tu-darmstadt.de
} %

\maketitle

\input{sections/abstract.tex}

\input{sections/introduction_new.tex}

\input{sections/background.tex}

\input{sections/model.tex}

\input{sections/overview.tex}

\input{sections/design.tex}

\input{sections/eval.tex}

\input{sections/discussion.tex}
\input{sections/comparison.tex}

\input{sections/relatedwork.tex}
\input{sections/conclusion.tex}

{
	\bibliographystyle{plain}
	\bibliography{bib}
}
\input{sections/appendix.tex}

\end{document}

%% file: sections/abstract.tex
\begin{abstract}
We address the challenging problem of efficient trust establishment in \emph{constrained networks}, 
i.e., networks that are composed of a large and dynamic set of (possibly heterogeneous) devices 
with limited bandwidth, connectivity, storage, and computational capabilities. 
Constrained networks are an integral part of many emerging application domains, 
from IoT meshes to satellite networks.
A particularly difficult challenge is how to enforce timely revocation of 
compromised or faulty devices. Unfortunately, current solutions and techniques cannot cope with idiosyncrasies of constrained networks, since they mandate frequent real-time communication 
with centralized entities, storage and maintenance of large amounts of revocation 
information, and incur considerable bandwidth overhead.

To address the shortcomings of existing solutions, we design \sysname,
a secure and efficient scheme for certificate validation that augments and benefits a
PKI for constrained networks. \sysname utilizes unique features of Sparse Merkle Trees (SMTs) to perform
lightweight revocation checks, while enabling collaborative operations among devices
to keep them up-to-date when connectivity to external authorities is limited.
\sysname can complement any PKI scheme to increase its flexibility and applicability, 
while ensuring fast dissemination of validation information independent of the network 
routing or topology. \sysname requires under 3KB storage per node covering $10^6$ 
certificates. 
We developed and deployed a prototype of \sysname on an in-orbit satellite
and our large-scale simulations demonstrate that \sysname decreases the number of requests for updates from external authorities by over 93\%, when nodes are intermittently connected.
\end{abstract}

%% file: sections/introduction_new.tex
\section{Introduction}
\label{sec:introduction}
Spurred by new and emerging applications---ranging from IoT to satellite networks---there has been a growing 
trend of interconnecting large numbers of heterogeneous resource-constrained devices in recent years.
In such settings, both devices and networking are constrained: devices have anemic computation 
and storage abilities, while networking is characterized
by limited bandwidth, low transmission range, dynamic topology, and more critically, by intermittent 
connectivity due to mobility and/or device hibernation. 
We use the term \emph{constrained networks} to describe such settings.

In particular, satellite networks constitute an emerging class of constrained networks~\cite{ipnsig}. Due to decreased 
satellite costs (e.g., \$22,000 for a \emph{CubeSat}, including launch~\cite{cubesatcost}) 
and their increased accessibility (e.g., AWS Ground Station service~\cite{awsgs}), 
the number of operational satellites has doubled to over 4,000 since 2019~\cite{satcount}.
Moreover, the trend towards constellations, i.e., deployment of a network of small satellites 
(instead of few large ones), will dramatically increase this growth in the years to come.
SpaceX's Starlink alone plans to deploy around 42,000 satellites~\cite{starlinkcount}.

However, small satellites have many constraints, such as strict power budgets, 
radiation-resistant hardware components with limited computing and storage capabilities~\cite{radcpus}, 
and physical transmission limitations, e.g., due to periodic line-of-sight blockage by planets and other 
celestial bodies. While satellite networks might seem to be an extreme example of constrained
networks, similar problems arise in terrestrial settings. For example, a number of mesh protocols 
have been designed to handle poor network conditions for low-power \ac{iot} 
devices~\cite{zigbee,zwave,blemesh,wifimesh}. Also, similar to line-of-sight disruptions in
satellite communication, home/office automation devices often hibernate to conserve power. 
Unattended outdoor IoT devices that use natural sources of power (solar, wind, etc.)
tend to hibernate. Moreover, mobile terrestrial devices can go out of range or encounter 
communication obstacles. All these conditions result in intermittent or unstable connectivity.

Efficient and secure trust establishment in constrained networks is essential. 
While small, static, homogeneous networks could rely on symmetric 
cryptography, large heterogeneous networks require scalable asymmetric cryptography.
\ac{pki} and public-key certificates are common tools deployed for establishing mutual trust between devices. 
However, timely certificate revocation of malfunctioning or compromised devices is critical 
for retaining trust in the whole network.  Since satellites also suffer from software 
bugs~\cite{satbugs} and can be subject to attacks~\cite{sathacks}, timely revocation is very
important. In fact, the Internet Engineering Task Force (IETF) already recognized the difficulty of revocation in a protocol slated 
for satellite networks~\cite{BPSec}.

There is also a large body of literature on \ac{pki} in distributed settings, such as 
observer-based approaches~\cite{ct,enhanced-ct,aki,arpki}, schemes that enable end-users to distributively check their certificates~\cite{coniks,dtki}, 
blockchain-based approaches~\cite{blockstack,ethiks}, and \ac{pki} that is specifically geared towards networks with delay tolerance~\cite{dtncrlrevoke,dtnsocial,dtnibc} as well as mobile 
ad-hoc networks~\cite{manetself,manetdistr,manetmoca,manetneighbor}.
Furthermore, recent efforts focus on \ac{pki} for \ac{iot}~\cite{colliotpki,pki4iot,iotbcpki}.

As discussed in \Cref{sec:related_work}, current techniques have some important 
shortcomings with regard to revocation checks in constrained networks:
First, they make strong assumptions about network connectivity or incur heavy communication overhead 
for the entire network. For instance, on-demand revocation checking, such as \ac{ocsp}~\cite{rfc6960}, 
requires a reliable connection and separate request for 
every certificate validation. However, a reliable connection to a central entity 
cannot be guaranteed in constrained networks. 
Second, storage and distribution of explicit revocation information, e.g., using 
\acp{crl}~\cite{rfc5280}, consumes high bandwidth and storage, including regular updates.
However, devices in a constrained networks can be highly limited in terms of storage 
and networking abilities, e.g., the popular Z-Wave low-power IoT technology has a 
bandwidth of 100Kbps at best~\cite{zwaveinfo}.
Even recent results that significantly reduce storage and update overheads of 
\acp{crl}~\cite{crlite, letsrevoke} are specifically designed for (generally reliable) 
Web-based revocation. Thus, they are poorly suited for an environment where devices 
frequently miss revocation updates. We discuss this in more detail in \Cref{sec:comparison}.

In summary, efficient certificate validation in constrained networks 
is still a challenging open problem that we aim to tackle in this paper.

\noindent\\
\textbf{Goals \& Contributions:}
We present \sysname, a novel certificate validation scheme for constrained networks.
\sysname provides lightweight certificate validation directly between devices, 
with minimal communication overhead, by defining operations that allow nodes 
to epidemically keep each other's revocation information up-to-date.
In a network with $10^6$ active certificates, \sysname requires under 
$3$KB of storage per device to allow all devices to mutually authenticate each other,
using widely available cryptographic primitives.
Furthermore, if devices miss revocation updates, \sysname reduces (by over $93\%$) 
the number of devices that need to request fresh validation information from the CA.
\\

\vspace{0.5em}\noindent
Our main contributions include:
\begin{compactitem}
	\item \sysname enables flexible and lightweight revocation checks in \ac{pki} schemes, especially for device-to-device trust establishment, thus enabling \ac{pki} in constrained networks.
	\item \sysname defines novel algorithms that utilize the deterministic structure of \acp{smt}, which allows devices to keep each other up-to-date.
	This eliminates the need for the vast majority of devices to contact the CA when updates were missed.
	\item \sysname introduces the \emph{Validation Forest} (VF) data structure for efficient exchange of validation information among devices, whenever they come in contact.
	VF allows for epidemic dissemination of validation information, without the need to consider 
	application-specific network aspects, such as the underlying topology or routing protocols.
	\item \sysname involves no on-demand requests, while requiring very little storage overhead for 
	devices. At the same time, it offers better security guarantees for constrained networks 
	than prior approaches.
	\item We evaluate \sysname's proof-of-concept implementation on the European Space Agency's OPS-SAT satellite. We then thoroughly evaluate \sysname in a large-scale simulation modeling a constrained network. We also open-source our prototype to the research community\footnote{\url{https://github.com/vcer4pki/VCER}}. 
\end{compactitem}

%% file: sections/background.tex
\section{Background}
\label{sec:background}
\acresetall

This section provides background on the data structure that is central for \sysname operation.

\subsection{Sparse Merkle Trees}
\label{sec:smt}
First proposed by Merkle~\cite{mht}, an \ac{mht} is an accumulator that efficiently represents a set 
of data elements and allows to construct \aclp{poi} for individual elements~\cite{mht}.
It requires a secure hash function, the digest of which is used to label each element contained in the set.
Knowing only the root hash of the tree, anyone can verify whether a given element (leaf) is part of the set 
by using a small set of hashes (co-path) corresponding to all sibling nodes on the path to root, 
i.e., a \ac{poi}, which is $O(\log n)$ where $n$ is the number of leaves.

An \ac{smt}~\cite{smt} is a type of \ac{mht} which contains all possible hash values, e.g., 
an \ac{smt} for SHA256 has $2^{256}$ leaves and depth of $255$. When inserting a new element into 
an \ac{smt}, the leaf representing a placeholder is replaced by the new element.
Because of the deterministic position of elements in the tree, removing elements is easy.
When an element is removed, the root hash changes and a previously valid \ac{poi} for the removed 
element becomes invalid.

While a complete \ac{smt} for a reasonable hash function (e.g., SHA256) is not computable, 
in practice, most leaves need not be assigned. Specifically, all ``empty'' leaves can be assigned 
$\texttt{H(}\varnothing\texttt{)}$. Going up such a tree, all hashes above any two empty leaves 
amount to: $\texttt{H(}\texttt{H(}\varnothing\texttt{)}\|\,\texttt{H(}\varnothing\texttt{)}\texttt{)}$, 
and so on. Therefore, using SHA256 as an example, we can construct all $256$ empty branch hashes 
for all depths in the tree stored in an \emph{EmptyHashesList}, which covers all hashes in the 
empty parts of the tree. Since most parts of this \ac{smt} are empty, we only need to 
consider assigned leaves to compute the root.

\begin{figure}[ht]
	\centering
	\includegraphics[width=0.7\columnwidth,trim=0cm 0.5cm 0cm 0cm, clip]{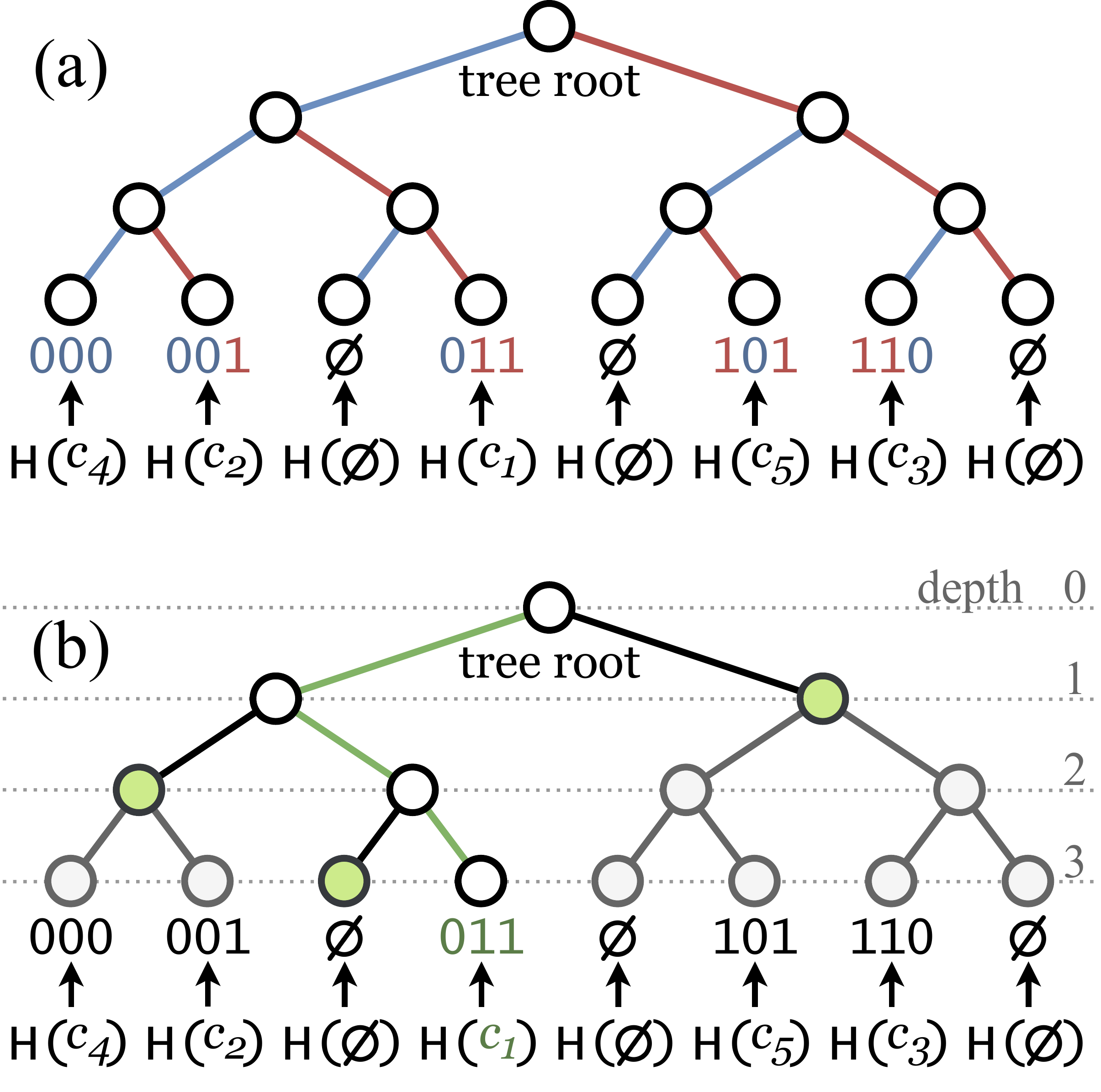} 
	\caption{\ac{smt} examples: (a) depicts how the position of hashes is determined and (b) shows a \ac{poi}
	and how depth numbering works.}
	\label{fig:smt}
\end{figure}

\Cref{fig:smt} (a) shows a sample \ac{smt} with the five elements $c_1,...,c_5$ for a hash function 
with a 3-bit digest. Branch colors show how the positioning of the elements works: blue represents 
an unset bit going left along the branch, and red is a set bit going right. For example, leaf $c_5$ 
has the hash digest $\texttt{H(}c_5\texttt{)}=101$, which represents its position in the tree. 
The three hash digests $010, 100$ and $111$ are not assigned in this example. \Cref{fig:smt} (b) shows 
the path for leaf $c_1$ as green lines in the tree. To construct its \ac{poi}, we need its sibling,
and all siblings of its ancestors, corresponding to the nodes in green. A \ac{poi} for $c_1$ is 
verified as follows:
\begin{compactenum}
    \item[\bf (i)]The verifier knows the (signed) root and the \ac{poi}:
    
{\footnotesize $
[\texttt{H(}\varnothing\texttt{)}, 
\texttt{H\textbf{(}}\texttt{H(}c_4\texttt{)}\|\,\texttt{H(}c_2\texttt{)}\texttt{\textbf{)}},
\texttt{H\textbf{(}}\texttt{H(}\texttt{H(}\varnothing\texttt{)}\|\,\texttt{H(}c_5\texttt{)}\texttt{)}\|\,
\texttt{H(}\texttt{H(}c_3\texttt{)}\|\,\texttt{H(}\varnothing\texttt{)}\texttt{)\textbf{)}}]
$}
    \item[\bf (ii)] The verifier computes the hash of $c_1$, combines it with the first element 
in the \ac{poi}, and hashes both. Next, the result is combined with the second element in the 
\ac{poi} and hashed. Finally, the last hash is combined with the third element in the \ac{poi} and hashed.
    \item[\bf (iii)] The verifier compares the last hash from (ii) with the root value. If they match, 
the \ac{poi} is valid.
\end{compactenum}
However, note that, since the sibling leaf of $c_1$ is empty, we can omit it. In a more complex example using 
SHA256, a \ac{poi} would need 256 elements, though most of them will be empty, which can be omitted with 
the help of the \emph{EmptyHashesList}. 
This requires an additional \emph{bitmap} the same size as the tree height, e.g., 256 bits for SHA256.
The bits in the bitmap at every tree level indicate whether an empty hash or an element of the \ac{poi} should be used (see \Cref{alg:calc_path_root} in \Cref{sec:indepth_algos} for details).
Because the digest of a strong cryptographic hash function is a pseudo-random value, 
the hashes of elements are uniformly distributed. Therefore, assigned leaves are evenly spread over the 
\ac{smt}, implying a \ac{poi} size of $\log n$ hashes, \emph{on average}.

%% file: sections/model.tex
\section{System Model}
\label{sec:system_model}
We consider three types of entities: \ca, \nodes, and cacher \nodes.
\nodes are devices that need to validate each other's public key certificates.
We use the term ``validation'' to focus on the revocation check, i.e., the chain-of-trust verification of certificates is implied.
Although a \node can compute hashes and verify signatures, it has limited processing power and storage.
A subset of \nodes play the role of cacher \nodes by storing additional information.
Communication is constrained due to low bandwidth, mobility, and intermittent connectivity that can 
result in frequent network partitions. \nodes exchange contact messages whenever they \emph{meet}.
\ca is the certificate-issuing authority. Communication with the \ca is particularly restricted.
We treat \ca as a single entity, albeit it 
might be distributed in practice.\footnote{\Cref{sec:discussion} discusses the setting with 
multiple {\ca}s.} Each \node knows \ca's certificate and trusts \ca's signatures. 
The number of certificates in the system is denoted by $n$.
For simplicity, we assume that each \node has exactly one certificate.
Further, we assume a coarse time synchronization among \nodes and \ca in the range of hours, e.g., to check certificate expiration.

\subsection{Adversary Model}
\label{sec:adversary_model}
We consider the Dolev-Yao model adversary (\adv) that can eavesdrop on, intercept, or inject any 
number of messages~\cite{dolevyao}. However, \adv is naturally bound to physical constraints 
of the network, such as not being able to reach a disconnected \node.
In particular, we assume that \adv acts locally, and cannot block all communications in the network at the same time.
We assume the \acl{smt} construction to be secure, i.e., collision-resistant.

\subsection{Requirements}
\label{sec:requirements}
As mentioned earlier, designing an efficient distributed certificate validation scheme for 
constrained networks is challenging, especially since nodes can miss updates, due to connectivity issues. 
We believe that an ideal scheme must satisfy the following requirements:
\begin{enumerate}[nolistsep, label=\textbf{R.\arabic*:}, ref={R.\arabic*}, leftmargin=0ex, 
labelsep=\widthof{~}, itemindent=\widthof{R.1:~\hspace{.125ex}}]
	\item \label{r:1}\textbf{Handle arbitrary delays:} Responses from a central entity (\ca) might be delayed or lost.
	\item \label{r:2}\textbf{Avoid single points of failure:} Although centralized systems are easy to set up, they fail when the central entity loses connectivity. Thus, we must avoid relying solely on central entities.
	\item \label{r:3}\textbf{Consistency:} While there is always current local state, validation decisions 
	must be derived from a common trust anchor to ensure consistency.
	\item \label{r:4}\textbf{Handle constrained devices:} A certificate validation scheme must have minimal impact on nodes' scarce resources.
	\item \label{r:5}\textbf{Timeliness:} Since freshest validation information is crucial for security, an
	ideal scheme must ensure regular updates and their fastest dissemination.
\end{enumerate}

%% file: sections/overview.tex
\section{\sysname Overview}
\label{sec:overview}
\sysname consists of three main components:
\begin{compactitem}
\item[\bf (i)] Certificate validation via an individual \ac{poi} for each \node.
\item[\bf (ii)] Efficient spreading of fresh validation information using 
    the \emph{Validation Forest} (\forest) data structure. \forest is the trust anchor 
    used to validate any \ac{poi}.
\item[\bf (iii)] Distributed repair whereby up-to-date \nodes directly help outdated
    \nodes to recover from missed updates.
\end{compactitem}

\begin{figure}[h]
	\centering
	\includegraphics[width=0.8\columnwidth]{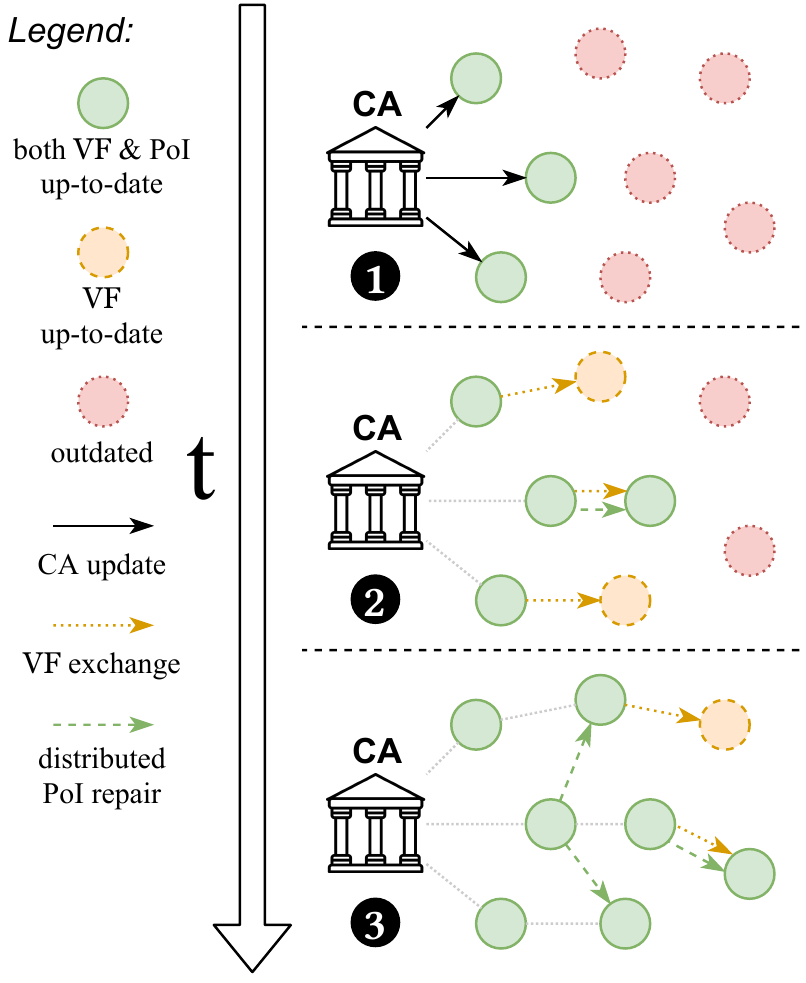}
	\caption{Example of \ca updates being spread in a distributed manner.}
	\label{fig:overview}
\end{figure}

\Cref{fig:overview} is a high-level overview of \sysname operation when there are \ca updates,
e.g., some certificates are newly revoked. After \ca updates the validation information
aggregated in \forest, all \nodes become outdated.
In step \circled{1}, \ca spreads its update information, with which each \node 
can update both its \forest and its \ac{poi}. However, in a constrained network, an
update would not reach all \nodes.  Specifically, the update reaches only the solid green \nodes 
in step \circled{1}, while all the rest miss this update (dotted red \nodes), e.g., 
by not being connected to the network or by simply hibernating.

After some time, in step \circled{2}, some outdated \nodes meet up-to-date \nodes and 
start to exchange information. Up-to-date \nodes update outdated nodes' \forest 
(dashed yellow arrows). With a fresh \forest, a \node can correctly validate other \nodes' 
certificates, e.g., reject newly revoked ones. This illustrates one key feature of \sysname:
any \node encountering an up-to-date node obtains the latest \forest on contact.
Thus, fresh validation information spreads very quickly and epidemically. 

An outdated \node's own \ac{poi} can become outdated if it misses some updates.
Although such a \node can still validate certificates of up-to-date nodes correctly, 
it cannot provide a proof for its own certificate validity to other \nodes (dashed yellow).
However, nodes with up-to-date \acp{poi} can help nodes with outdated ones to update their \ac{poi} 
via distributed repair (green dashed arrows). For this, we exploit the deterministic 
structure of \acp{smt} to design operations for distributed repair, as shown in \Cref{sec:distr_repair}.
In step \circled{2} this succeeds for the node in the center.
However, after some time passes in step \circled{3}, nodes continue to encounter others, 
increasing their chance for distributed repair to succeed.
Eventually, most nodes would successfully repair their outdated \ac{poi}, with 
only a few having the need to directly request a fresh one from the \ca.
We demonstrate this in \Cref{sec:eval_sim}.

%% file: sections/design.tex
\section{\sysname Certificate Validation Scheme}
\label{sec:design}
This section describes all components of \sysname.
For certificate validation, \nodes use \aclp{poi} to verify that a \node's certificate is valid.
In a simplified example, the \ca can build an \ac{smt} with all the active certificates' hashes 
and \nodes only need to know the root hash of the \ac{smt} to verify any \ac{poi}.
Each \node then stores the \ac{poi} for its own certificate, and becomes capable of proving 
validity of its certificate to others.
This is done by computing the \ac{poi}'s root hash (see \Cref{alg:calc_path_root} in 
\Cref{sec:indepth_algos}) and if it matches the root hash given by \ca, the certificate is valid.
Thus, \nodes only need to store the respective tree root and their own \ac{poi}.

\begin{figure}[h]
	\centering
	\includegraphics[width=0.75\columnwidth]{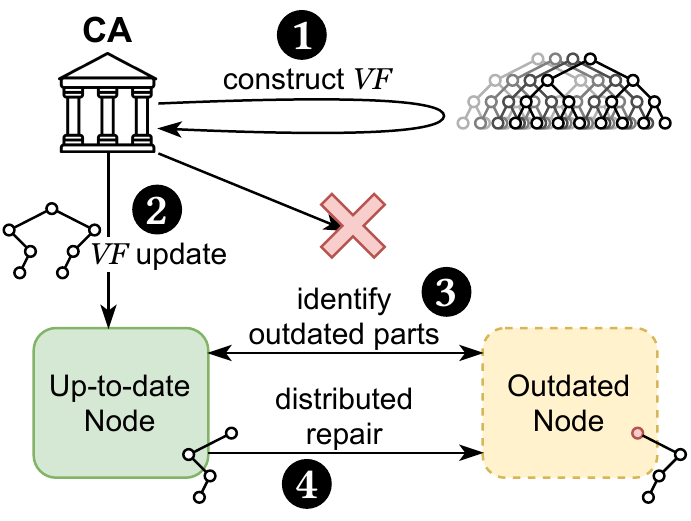}
	\caption{The individual operations of \sysname.}
	\label{fig:flowchart}
\end{figure}

\Cref{fig:flowchart} shows individual operations in \sysname.
In step \circled{1}, the \ca constructs the Validation Forest \forest, as described in \Cref{sec:forest}.
Instead of using a single \ac{smt}, \forest is a data structure used by \nodes to efficiently keep their certificate validation information up-to-date.
Upon any changes in \forest, e.g., revocations, the \ca constructs updates that are processed by the \nodes in step \circled{2}, which is described in \Cref{sec:ca_updates}.
As some \nodes may miss these updates, in step \circled{3}, \nodes exchange information to keep each other updated and identify which parts of \forest are outdated (see \Cref{sec:prime_exchange}).
Finally in step \circled{4}, \nodes repair each other's \acp{poi} when \ca updates are missed, as presented in \Cref{sec:distr_repair}.
We accompany our description with a running example that uses practical parameters.
The evaluation in \Cref{sec:eval_perf} is also based on this example.
In terms of cryptographic primitives, the example uses SHA256 
and ECDSA signatures based on the \emph{secp256r1} curve.

\subsection{Validation Forest}
\label{sec:forest}
The Validation Forest \forest is the core data structure for validating certificates in \sysname.
As long as a \node's \forest is up-to-date, it can correctly validate another \node's certificate 
with its \ac{poi}. \forest has three main parts. 
(i) A number of \ac{smt} roots for \ac{poi} validation.
(ii) \primename \primem is a small data structure used for efficient exchange 
between \nodes to keep their \forest up-to-date.
(iii) \ca's signature of \primem (\primesig), which is the trust anchor for all operations in \sysname.
add phrasing that VF represents active cert set

The number of trees in \forest is the number of epochs it models.
Certificates have a defined maximum lifetime and this lifetime is split into \epochs epochs.
A certificate's expiration date is then used to assign it to the respective epoch tree 
\epochtree, represented by the epoch root \epochroot stored in \forest.
Each node stores the \forest including all \epochroot and keeps it up-to-date.
A \node's own proof is valid for a single \epochroot, which can be inferred from the certificate's 
expiration date. 
This stabilizes the individual \acp{smt}, significantly reducing the number of potentially outdated \acp{poi} on any updates by \ca. 
After an epoch has passed, a \emph{forest prune} occurs, i.e., the oldest tree is pruned, 
as it contains only expired certificates.
This means that \ca and \nodes can discard the oldest \epochtree and the corresponding \epochroot.

For our accompanying example, we define certificates with a maximum lifetime of around one year, 
split into weeks as epochs. Thus, \epochs=52, meaning that the \forest stores a maximum of
52 \epochroot, each representing certificates for the corresponding week of expiration. 
This requires each node to store up to 52 hashes for the tree roots, resulting in \textasciitilde 
1.7kB of storage. Assuming that changes occur mostly in the newest 
epoch, e.g., when new certificates are issued, other epoch trees are left 
untouched, including their \epochroot. Thus, all current \acp{poi} depend on them.
This amount of overhead fulfills requirement \ref{r:4} for low-storage devices.
Other certificate lifetime configurations are possible; however, the primary overhead factor is the revocation update frequency, as we show in \Cref{sec:eval_sim}.

\subsubsection{\primename}
\label{sec:prime}
The \primename \primem is used to exchange key information by \nodes about the current \forest state 
to help keep it up-to-date among each other. It is designed to be lightweight to allow its 
inclusion in contact messages \nodes exchange when they meet (cf. \Cref{sec:prime_exchange}).
\primem includes \primename root \primeroot, timestamp \primets, and checksums \primeparity
to identify outdated tree roots. Furthermore, \primem is signed by \ca. \primename signature \primesig 
serves as the trust anchor for \nodes to confirm outcomes of all operations (requirement \ref{r:3}).

\primeroot is computed by simply concatenating all tree roots in \forest and hashing 
resulting information. Thus, any change in any tree would result in a different \primeroot, 
and \nodes can use it to check if any parts of their \forest is outdated.
To distinguish among multiple \primem in terms of freshness, each contains a timestamp \primets.

\primem also contains the checksums \primeparity for all \epochroot in \forest to efficiently identify outdated \epochroot.
This avoids having a \node send all tree roots to its peer with an outdated \forest every time.
\primeparity is split into two types: 
main and aggregated. Each checksum type has a configurable size.
Main checksums are applied directly to tree roots representing newest epochs in \forest.
Aggregated checksums are applied on a number of concatenated tree roots  
that come after the ones covered by main checksums. The number that is aggregated into a 
single checksum is configurable in \sysname.
Note, while we use the term ``checksum'', it is sufficient to simply use some bytes of the hash, as the \ac{smt} protects against collisions.
This way, tree roots that are expected to change more often than others are covered by their own 
checksum, e.g., newly issued certificates are inserted into the newest tree.
In contrast, \epochtree that are expected to be more stable get aggregated checksums.

In our example, we use 2 main checksums and aggregate 10 tree roots per aggregated checksum, 
resulting in 5 aggregated checksums for a total of 7. For \primets, we use a 4 Bytes-long 
UNIX-timestamp. Using 2 Bytes per checksum, this results in \primem~=~50 Bytes (\primeroot~=~32B, 
\primeparity~=~14B, \primets~=~4B). Also, \ca's signature \primesig over \primem is 
64 Bytes. This amount of overhead fulfills requirement \ref{r:4} for low-bandwidth devices.

\subsection{CA Updates}
\label{sec:ca_updates}
When any changes occur, \ca updates the respective \epochtree resulting in changes for both 
\forest and \acp{poi}. Thus, \ca needs to distribute updates for \nodes to be up-to-date.
To keep the \nodes' \forest updated, \ca simply needs to distribute new \primem, including new
\primesig and \epochroot for affected epochs.
Also, if there are no updates for a while, \ca can regularly send out current \primem with new \primets. 
This way, \nodes eventually realize that they are outdated, e.g., even when disconnected for a while.

While this keeps \nodes' validation information updated, any \ac{poi} found in epochs affected by 
an update becomes invalid, and respective \nodes can no longer provide a valid proof for their certificates.
Instead of \ca individually distributing updated \acp{poi} to all affected \nodes, it constructs universal 
updates for all \nodes. This is done by distributing \acp{poi} for all updated certificates, 
including revoked ones. Due to the deterministic order of elements in \ac{smt}, \nodes can 
process the update \acp{poi} affecting their own epoch to update their \ac{poi}.
Afterwards, the update \acp{poi} are discarded.
Furthermore, when an update contains many \acp{poi}, there are likely many redundant \ac{poi} 
elements that can be aggregated to reduce update size.

\begin{algorithm}[h]
	\caption{\texttt{update\_poi\_with\_poi} function for updating a proof of inclusion regarding an up-to-date proof of inclusion. \\
	{\small \emph{int}$\langle p \rangle$ accesses the $p$-th bit  of \emph{int} from the right,
	\textbar\texttt{H}\textbar\ is the bit length of the hash digest and \textbar$l$\textbar\ is the number of elements in list $l$. The variables ending in \emph{path} are lists, \emph{hash} and \emph{bitmap} variables are integers that fit \textbar\texttt{H}\textbar.
	}}
\label{alg:update_poi_with_poi}
	\small
	\raggedright
	\begin{algorithmic}[1]
		\Require \emph{my\_leaf\_hash}, \emph{my\_path}, \emph{my\_path\_bitmap}, \emph{new\_leaf\_hash}, \emph{new\_path}, \emph{new\_path\_bitmap}
		\Ensure \emph{my\_path}, \emph{my\_path\_bitmap}
		
		\State \emph{xor\_leaves} $\gets$ \emph{my\_leaf\_hash} $\oplus$ \emph{new\_leaf\_hash}
		\label{alg:update_poi_with_poi:xor1}
		\State \emph{target\_pos} $\gets$ {\footnotesize position of left-most set bit in} \emph{xor\_leaves} {\footnotesize from the left}%
		\label{alg:update_poi_with_poi:xor2}
		
		\State \emph{is\_update} $\gets$ \texttt{False}
		\label{alg:update_poi_with_poi:path1}
		\State \emph{path\_pos} $\gets$ 0
		\For{$i \gets 0$ \textbf{to} (\emph{target\_pos} $+$ 1)}
		\If{\emph{my\_path\_bitmap}$\langle$\textbar\texttt{H}\textbar\ $-$ 1 $-$ $i$$\rangle$ $=$ True}
		\If{$i = $ \emph{target\_pos}}
		\State \emph{path\_pos} $\gets$ \emph{path\_pos} + 1
		\State \emph{is\_update} $\gets$ True
		\Else
		\State \emph{path\_pos} $\gets$ \emph{path\_pos} + 1
		\State \emph{my\_path}[\textbar\emph{my\_path}\textbar\ $-$ \emph{path\_pos}]
		\Statex \hspace{10em}
		$\gets$ \emph{new\_path}[\textbar\emph{new\_path}\textbar\ $-$ \emph{path\_pos}]
		\EndIf
		\EndIf
		\EndFor
		\label{alg:update_poi_with_poi:path2}
		\Statex
		\Comment{{\footnotesize loop finds update-bit-position and if it affects existing hash in path}}
		
		\State \emph{update\_hash} $\gets$ \texttt{calc\_path\_root(}\emph{new\_leaf\_hash},
		\Statex \hspace{6em}
		\emph{new\_path}, \emph{new\_path\_bitmap},	(\emph{target\_pos} $+$ 1)\texttt{)}%
		\label{alg:update_poi_with_poi:hash}
		\If{\emph{is\_update}}
		\State \emph{my\_path}[\textbar\emph{my\_path}\textbar\ $-$ \emph{path\_pos}] $\gets$ \emph{update\_hash}
		\Statex
		\Comment{{\footnotesize replace exiting hash in path}}
		\Else
		\State \emph{my\_path}\texttt{.insert(}(\textbar\emph{my\_path}\textbar\ $-$ \emph{path\_pos}), \emph{update\_hash}\texttt{)}
		\State \emph{my\_path\_bitmap}$\langle$\textbar\texttt{H}\textbar\ $-$ 1 $-$ \emph{target\_pos}$\rangle$ $\gets$ True
		\Statex
		\Comment{{\footnotesize insert new hash in path \& set respective bit in bitmap}}
		\EndIf
		
		\State \textbf{return} \emph{my\_path}, \emph{my\_path\_bitmap}
	\end{algorithmic}
\end{algorithm}

\Cref{alg:update_poi_with_poi} shows how \nodes process update \acp{poi} provided by \ca.
The operation takes an up-to-date \ac{poi} and updates an outdated \ac{poi} found in the same epoch.
This is then done for all \acp{poi} in the update.
The path-bitmaps work as described in \Cref{sec:smt}.
First, the hash of the outdated certificate is XOR-ed 
with the updated hash and the position of the left-most set bit shows where both \acp{poi} split in the tree.
Afterwards, the algorithm checks if the outdated \ac{poi} already has an element at the split position, 
triggering an overwrite of this existing element in the \ac{poi}.
Otherwise, the outdated \ac{poi} needs an additional element at the split position, including a set 
bit in the bitmap. The hash is computed by calling \texttt{calc\_path\_root} (see \Cref{alg:calc_path_root} 
in \Cref{sec:indepth_algos}), the same algorithm used to validate \acp{poi}, except the optional 
last parameter indicates the need to stop at the specified depth, instead of the root hash.

Before the split position is reached, every \ac{poi} element is updated along the way.
This effectively allows to blindly apply updates, meaning a \node does not need to worry 
about the order of applying updates.
As long as the \acp{poi} are up-to-date, \Cref{alg:update_poi_with_poi} does not perform 
any destructive updates. Afterwards, the executing \node can simply check if the resulting 
\ac{poi} is valid. Under the right circumstances, this may even cover for previously missed updates.
We take advantage of this aspect for the distributed repair, described in \Cref{sec:directrepair}.

\paragraph{Epoch Change.}
Whenever a new epoch starts, there might be many newly issued certificates.
On one hand, the aforementioned forest prune occurs (see \Cref{sec:forest}), 
i.e., many certificates will expire. At this point,
many \nodes will get a new certificate.
On the other hand, new \nodes may join the network, likewise with new certificates.
In \sysname, the issuing of new certificates should be aggregated until an epoch change occurs to increase efficiency.
This way, instead of constructing an update bundling many new \acp{poi}, \ca can distribute all the certificates hashes as the update, i.e., all epoch tree leaves.
This reduces the update size for the epoch change and all \nodes that have been issued a new certificate can construct their \ac{poi} themselves.
Further, this update only needs to be sent to \nodes that are affected by the epoch change.
The rest of \nodes only need the new epoch root for their \forest.
Additionally, in case an epoch does not contain any revoked certificates, the respective \epochroot can be set to $\texttt{H(}\varnothing\texttt{)}$, indicating to the network that any certificates in this epoch are not revoked.

\subsection{\primename Exchange}
\label{sec:prime_exchange}

When a \node missed any updates from \ca, its \forest will be outdated, and thus it will not have fresh certificate validation information.
A key aspect of \sysname is for \nodes to be able to efficiently keep each other up-to-date.
For this, \nodes exchange their \primename \primem (cf. \Cref{sec:prime}) with the contact message when they meet.
Additionally, \nodes will add information about which epoch they belong to and if they have any caches ready.
Both are important for the distributed repair, explained in \Cref{sec:distr_repair}, and in our accompanying example 1 Byte is sufficient (6 bits for the epoch and 2 bits as flags for caches).

\nodes only need an up-to-date \forest to be able to correctly validate \acp{poi} for any certificate.
This ensures that the current validation information spreads as quickly as possible throughout the network, without the need to consider any transmission aspects, such as acknowledgments from each \node to \ca to all \nodes are up-to-date.
For example, if a certificate was revoked and \ca sends out an update, even a \node missing this update will meet an up-to-date node eventually and after the \primem exchange, inherently know about the revocation, i.e., reject the revoked certificate's \ac{poi}.
Thus, this meets the requirement \ref{r:2} and \ref{r:5} for validating certificates.

\begin{figure}[p]
	\centering
	\includegraphics[width=0.96\columnwidth]{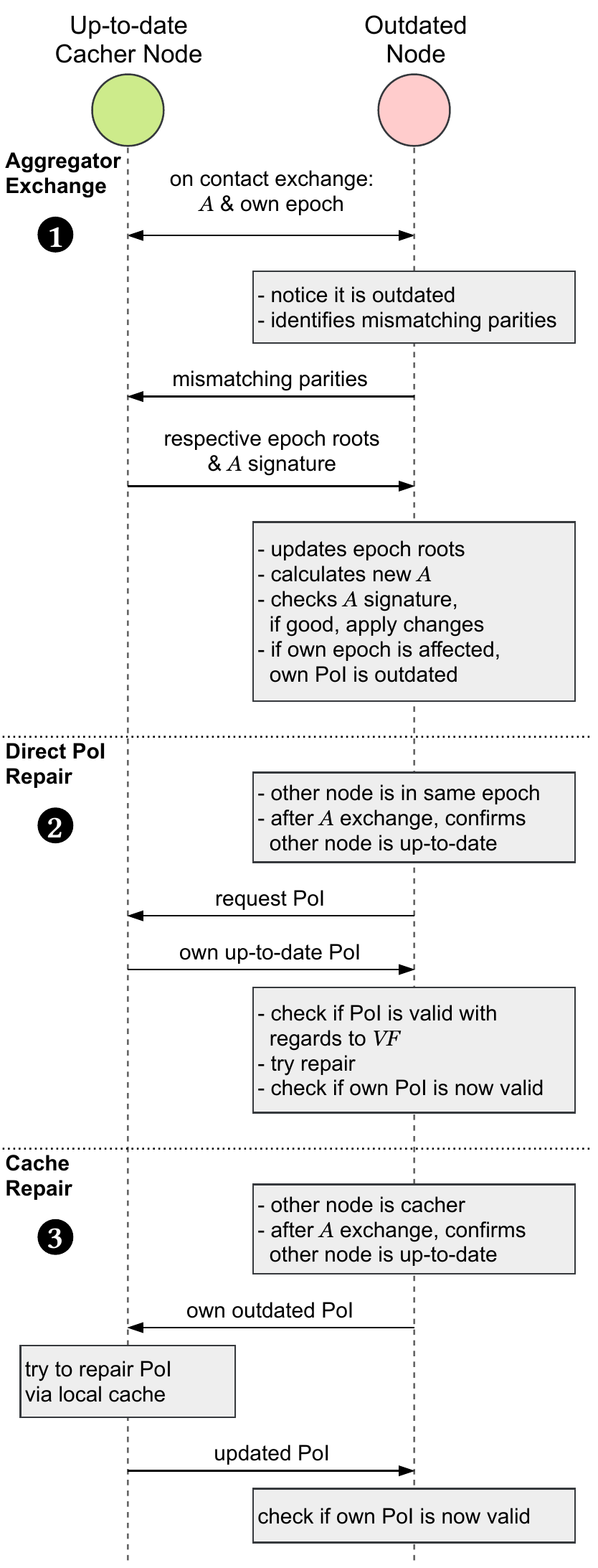}
	\caption{Exchange of \nodes for distributed repair for the three main operations.}
	\label{fig:exchange}
\end{figure}

\Cref{fig:exchange} \circled{1} depicts such an exchange in detail.
After both exchanged their \primem, the outdated \node on the right will see a new \primename root \primeroot as well as a newer timestamp \primets, and realize its \forest is outdated.
To identify which trees are actually affected by the change, the outdated \node will check the parities \primeparity and send the up-to-date \node the respective outdated checksum identifiers.
The up-to-date \node will answer with the respective \epochroot and add \primesig.
The outdated \node is then able to update all of the tree roots in its \forest, re-calculate its \primem, and finally check if \primesig is valid.
If one of the actually changed \epochroot is found in the same epoch as the \node's own certificate, it can imply that its \ac{poi} is outdated.
There is also the case that \epochroot has changed; yet, the corresponding \primeparity resulted in the same as before.
In this unlikely case, the outdated \node has to request all \epochroot, i.e., \forest.

In our example, let us assume there has been a change in the second newest epoch (covered by a main checksum) and 26th epoch in the middle. 
Thus, the outdated \node will request the second and fifth checksum.
The up-to-date \node will respond with the 11 respective \epochroot, which allows the outdated \node to update its \forest.
This entire exchange will require around 470 Bytes of communication overhead (\primem + \primesig + 11 $\cdot$ \epochroot) directly between the two \nodes.

\subsection{Distributed Repair}
\label{sec:distr_repair}
While the \primem exchange ensures that \nodes update their certificate validation information as fast as possible, missed updates may also lead to a \node's own \ac{poi} to be outdated.
However, a key goal of \sysname is to avoid having the outdated \node to contact \ca and request a fresh \ac{poi}.
Otherwise, many \nodes individually requesting fresh \acp{poi} at a similar time leads to a significant overhead for the entire network.
Thus, in this section we will present ways how an up-to-date \node can directly help an outdated \node to repair its \ac{poi}.
After an outdated \node updated its \primem, it will realize its \ac{poi} is not valid and can start requesting repair information from up-to-date \nodes it meets.
In the following, we will present two different approaches for this.
One is directly leveraging up-to-date \acp{poi} from other \nodes.
The other approach introduces a cache, stored and maintained by a share of \nodes in the network for increased efficacy of distributed repairs.
With these operations, \sysname meets requirement \ref{r:2} for a \node's own validation proof.

\subsubsection{Direct PoI Repair}
\label{sec:directrepair}
With the direct \ac{poi} repair strategy, an outdated \node can collect up-to-date \acp{poi} from other \nodes it meets to potentially repair its outdated \ac{poi}.
\Cref{fig:exchange} \circled{2} depicts how such an exchange proceeds.
After the initial \primem exchange on contact, the outdated \node knows if the other is up-to-date and if its certificate is found in the same epoch.
If both are true, the outdated \node will request the other's \ac{poi}, check if it is actually valid regarding the current \epochroot, and if so, use it for repair.

The outdated \node leverages \Cref{alg:update_poi_with_poi} for this, as mentioned in \Cref{sec:ca_updates}.
The operation replaces all applicable elements in the outdated \ac{poi} with the elements in the up-to-date \ac{poi}.
If the position of the up-to-date \node's leaf is favorable, it will update one or more elements in the outdated \ac{poi}.
In an unfavorable case, elements will simply remain unchanged.
This process can be repeated with other up-to-date \acp{poi} in the same epoch, until the \node's own \ac{poi} is valid regarding the current \epochroot.
This gets more difficult the more updates a \node missed for its epoch, which we evaluate in \Cref{sec:directanalysis}.
As \Cref{alg:update_poi_with_poi} works by blindly replacing elements in the \ac{poi}, the outdated \node must first verify the \ac{poi} used for the repair is actually valid for the current \forest.

\subsubsection{Level-Cache Repair}
\label{sec:lcrepair}

For the Level-Cache (\emph{LC}) strategy, a share of \nodes with larger storage capacity,  called \emph{cacher}, may additionally keep all hashes of each \epochtree on a specified depth, i.e., the cache level (\emph{clvl}).
This results in a storage overhead of 2\textsuperscript{\emph{clvl}} hashes per epoch.
As mentioned in \Cref{sec:smt}, the leaves of an \epochtree are uniformly distributed.
Due to this fact, the higher depth levels of the \ac{smt} are likely to be assigned before the lower ones.
If enough elements were inserted, the first few depth levels of a \ac{smt} will form a fully filled sub-tree.
Further, most updates in its \ac{poi} from the perspective of one \node will likely be in this sub-tree.

To clarify this phenomenon, consider the example of a \ac{smt} with one leaf that is only zeros.
Any new leaf that is inserted will go a different path at one point in the tree, which creates the need for an additional element in the \acp{poi} for both leaves.
The chances that any new leaf will branch off to the second half of the tree, i.e., the leaf having a set bit on the most-left position, are 50\%.
The likelihood of branching off on the second depth level is 25\%, and halves for any further depth level.
Thus, when inserting many leaves, the first depth levels in the \ac{smt} will likely branch off first.
This implicitly means, that given an up-to-date \emph{LC}, outdated \nodes are likely able to repair their own \ac{poi} with it.
The exact probabilities of this are discussed in \Cref{sec:eval_distr_repair}.
Furthermore, \nodes may use \emph{LC}s with different \emph{clvl}s, e.g, more resourceful \nodes may keep a larger cache and more limited \nodes a smaller one, if any at all.
To avoid having to send the entire \emph{LC} to an outdated \node, it will instead send its outdated \ac{poi} to the cacher \node, as shown in \Cref{fig:exchange} \circled{3}.

In our example, we chose to equip a share of \nodes with a \emph{LC} for \emph{clvl} = 7.
This means that these cacher \nodes will store an additional 4KB per epoch, and thus 208KB in total.
In \Cref{sec:discussion}, we will discuss possible alternative strategies; yet, for our purposes we set all cachers to store a \emph{LC} with the same \emph{clvl} for all epochs.

\begin{algorithm}[h]
	\caption{\texttt{update\_lvl\_cache\_with\_poi} for updating the level-cache regarding a new proof of inclusion}
	\label{alg:update_lvl_cache}
	\small
	\raggedright
	\begin{algorithmic}[1]
		\Require \emph{LC}, \emph{clvl}, \emph{new\_leaf\_hash}, \emph{new\_path}, \emph{new\_path\_bitmap}
		\Ensure \emph{LC}
		
		\State \emph{delete\_bits} $\gets$ $2$\textsuperscript{(\textbar\texttt{H}\textbar\ $-$ \emph{clvl})} $-$ 1
		\label{alg:update_lvl_cache:part1}
		\State \emph{part\_no} $\gets$ \emph{new\_leaf\_hash} $\&$ $\sim$\emph{delete\_bits}
		\State \emph{part\_no} $\gets$ \emph{part\_no} $\gg$ (\textbar\texttt{H}\textbar\ $-$ \emph{clvl})
		\label{alg:update_lvl_cache:part2}
		\Statex
		\Comment{{\footnotesize take \emph{clvl} left-most bits that define position in \emph{LC}}}
		\State \emph{update\_hash} $\gets$ \texttt{calc\_path\_root(}\emph{new\_leaf\_hash}, 
		\Statex \hspace{11em}
		\emph{new\_path}, \emph{new\_path\_bitmap}, \emph{clvl}\texttt{)}
		\Statex
		\Comment{{\footnotesize calculate path's root, but stop at depth \emph{clvl}}}
		\State \emph{LC}[\emph{part\_no}] $\gets$ \emph{update\_hash}
		
		\State \textbf{return} \emph{LC}
	\end{algorithmic}
\end{algorithm}

However, the cacher needs to keep its \emph{LC} up-to-date as well.
The construction of the \emph{LC} is described in \Cref{sec:lc_construction}.
The cacher uses \Cref{alg:update_lvl_cache} to process \ca update \acp{poi} for one epoch.
First, it extracts the correct position in \emph{LC} to be updated by the new \ac{poi}.
Afterwards, it uses the \ac{poi} root hash calculation, with the only difference that it stops at \emph{clvl} and inserts the resulting hash at the respective position of \emph{LC}.
When the cacher itself misses \ca updates, it may also meet other cacher \nodes and request up-to-date \emph{LC}s from them.

\begin{algorithm}[ht]
	\caption{\texttt{update\_poi\_with\_lvl\_cache} for updating a proof of inclusion with a given level-cache}
	\label{alg:update_poi_with_lvl_cache}
	\small
	\raggedright
	\begin{algorithmic}[1]
		\Require  \emph{my\_leaf\_hash}, \emph{my\_path}, \emph{LC}, \emph{clvl}
		\Ensure \emph{my\_path}
		
		\State \emph{delete\_bitmap} $\gets$ $2$\textsuperscript{(\textbar\texttt{H}\textbar\ $-$ \emph{clvl})} $-$ 1
		\State \emph{part\_no} $\gets$ \emph{my\_leaf\_hash} $\&$ $\sim$\emph{delete\_bitmap}
		\State \emph{part\_no} $\gets$ \emph{part\_no} $\gg$ (\textbar\texttt{H}\textbar\ $-$ \emph{clvl})
		
		\State \emph{part\_neg} $\gets$ $\sim$\emph{part\_no}
		\label{alg:update_poi_with_lvl_cache:neg}
		\Statex
		\Comment{{\footnotesize negated bits of \emph{part\_no} define its hash-neighbor in \emph{LC}}}
		\For{$i\gets 0$ \textbf{to} \emph{clvl}}
		\State \emph{new\_hash} $\gets$ \texttt{calc\_pos\_in\_LC(}\emph{part\_neg}, ($i+1$), 
		\emph{LC}\texttt{)} 
		\Statex 
		\Comment{{\footnotesize constructs a hash at position \emph{part\_neg} at depth ($i+1$) with the \emph{LC}}}
		\State \emph{my\_path}[\textbar\emph{my\_path}\textbar\ $-$ 1 $-$ $i$] $\gets$ \emph{new\_hash}
		\State \emph{part\_neg}$\langle$\emph{clvl} $-$ 1 $-$ $i$$\rangle$ $\gets$ $\sim$\emph{part\_neg}$\langle$\emph{clvl} $-$ 1 $-$ $i$$\rangle$
		\Statex
		\Comment{{\footnotesize flip bit to get hash-neighbor for next depth}}
		\EndFor
		
		\State \textbf{return} \emph{my\_path}
	\end{algorithmic}
\end{algorithm}

To update a \ac{poi} with a \emph{LC}, \Cref{alg:update_poi_with_lvl_cache} is used.
Like in \Cref{alg:update_lvl_cache}, the first three lines construct the position of the targeted cache element in \emph{LC}.
Then, for all \emph{clvl}, the respective neighborhood hash of the outdated \ac{poi} is constructed with the help of \texttt{calc\_pos\_in\_LC}.
This function simply takes a position and depth, which are used to construct a hash further up the tree in \emph{LC}.
The resulting hash is then set in the correct position in the outdated \ac{poi}.
This process is repeated for all depth levels in \emph{LC}, as the lower depth \ac{poi} elements can be updated as well.
This operation blindly replaces \ac{poi} elements and may not entirely repair the leaf's \ac{poi}.
However, if it cannot entirely repair the \ac{poi} it will still repair elements that \emph{LC} provides and at least assists in the overall repair process of the outdated \node.
In \Cref{sec:eval_distr_repair} we show that \emph{LC} can repair outdated \acp{poi} with high probability, given appropriately chosen parameters.

%% file: sections/eval.tex
\section{Evaluation}
\label{sec:eval}
In this section, we evaluate \sysname regarding multiple aspects. 
First, we evaluate its security, followed by an analysis on the success probabilities of the distributed repair approaches we introduced.
Finally, we will consider \sysname's performance regarding run-time overhead and large-scale networks.

\subsection{Security}
\label{sec:eval_security}

As \sysname provides the means to validate certificates, the adversary \adv aims to convince \nodes that (i) a revoked or forged certificate is valid, or (ii) a valid certificate is revoked.
In the following, we will explore different strategies \adv may use to achieve this goal and explain how \sysname prevents their success.

\paragraph{Manipulating Updates.}
To achieve either (i) or (ii), \adv can try to counterfeit updates by disseminating a new \epochroot that reflects the false state of the targeted certificate, i.e., a valid \ac{poi}.
However, a false \epochroot in \forest would also result in a different \primeroot.
As \primem is signed by \ca via \primesig, \nodes with an up-to-date \primem will discard the counterfeit update as invalid.

\paragraph{Blocking Updates.}
\adv can isolate a \node preventing \ca updates to reach it or delay update messages such that they receive them after the validation.
This way, a recently compromised certificate 
would still be validated regarding the outdated \forest, making \adv achieve (i).
This requires \adv to block any contact of the outdated \node with up-to-date \nodes, to prevent them to update their \forest.
This gets increasingly more difficult with increased network connectivity and more opportunities to meet nodes.
Nevertheless, as \ca regularly sends out an updated \primem, even isolated \nodes will eventually consider their current \forest to be outdated, and thus reject any \acp{poi}.
This limits the vulnerability window for \adv, similar to \acp{crl} or certificates with a limited lifetime, as considered in related works (cf. \Cref{sec:related_work}).
However, we consider completely isolating \nodes as non-trivial, and thus \sysname provides better security than schemes relying on a more centralized distribution of updates, as \adv needs to prevent a \node to communicate with more entities.

\paragraph{Denial-of-Service.}
Instead of preventing updates from reaching an individual \node, a powerful \adv could also perform a DoS attack on the entire network to prevent it from receiving updates to achieve (i).
Expecting a regular update of \primem by \ca, all \nodes will eventually consider their \forest to be stale.
Again, the vulnerability window is similar to other revocation checks.

\paragraph{Destructive Repair.}
To achieve (ii), \adv can target a \node with an outdated \ac{poi} and send it false repair information to prevent it from obtaining a valid proof for its certificate after an applicable \ca update.
As described in \Cref{sec:distr_repair} an outdated \node ensures the correctness of repair information regarding its up-to-date \forest before applying the repair.
This way, any invalid repair information is detected.
Further, meeting any benign \nodes may result in a repaired \ac{poi} anyway and for \adv to prevent this, requires it to isolate the \node.
Finally, a \node with an outdated \ac{poi} will give up on repairs eventually and request its valid \ac{poi} directly from \ca.

\subsection{Distributed Repair Analysis}
\label{sec:eval_distr_repair}
In this section, we evaluate the effectiveness of distributed repair approaches.
Note that the following results are for a single \ac{smt}; yet, \forest will consist of a tree per epoch.
For example, assuming individual \ac{poi} updates are evenly distributed among all epochs, the number of total missed updates needs to be divided by the number of epochs.
For this, we ran simulations on a pre-generated \ac{smt} with 100,000 random leaves and averaged the results over 10,000 runs.

\paragraph{Direct PoI Repair Analysis}
\label{sec:directanalysis}
We now present simulation results regarding the repair of an outdated \ac{poi} by using random up-to-date \acp{poi}, as described in \Cref{sec:directrepair}.
From the perspective of an outdated \node, the key driver for success is the distance of the missed updated leaves from the \node's own leaf in the \ac{smt}.
The closer any missed updated leaf is, the fewer \acp{poi} overall can help with the repair.
The more updates are missed, the higher the probability that one of them is unfavorable for an outdated \node.

\begin{figure}[h]
	\centering
	\includegraphics[width=1\columnwidth,trim=0.25cm 0.25cm 0.25cm 0.25cm, clip]{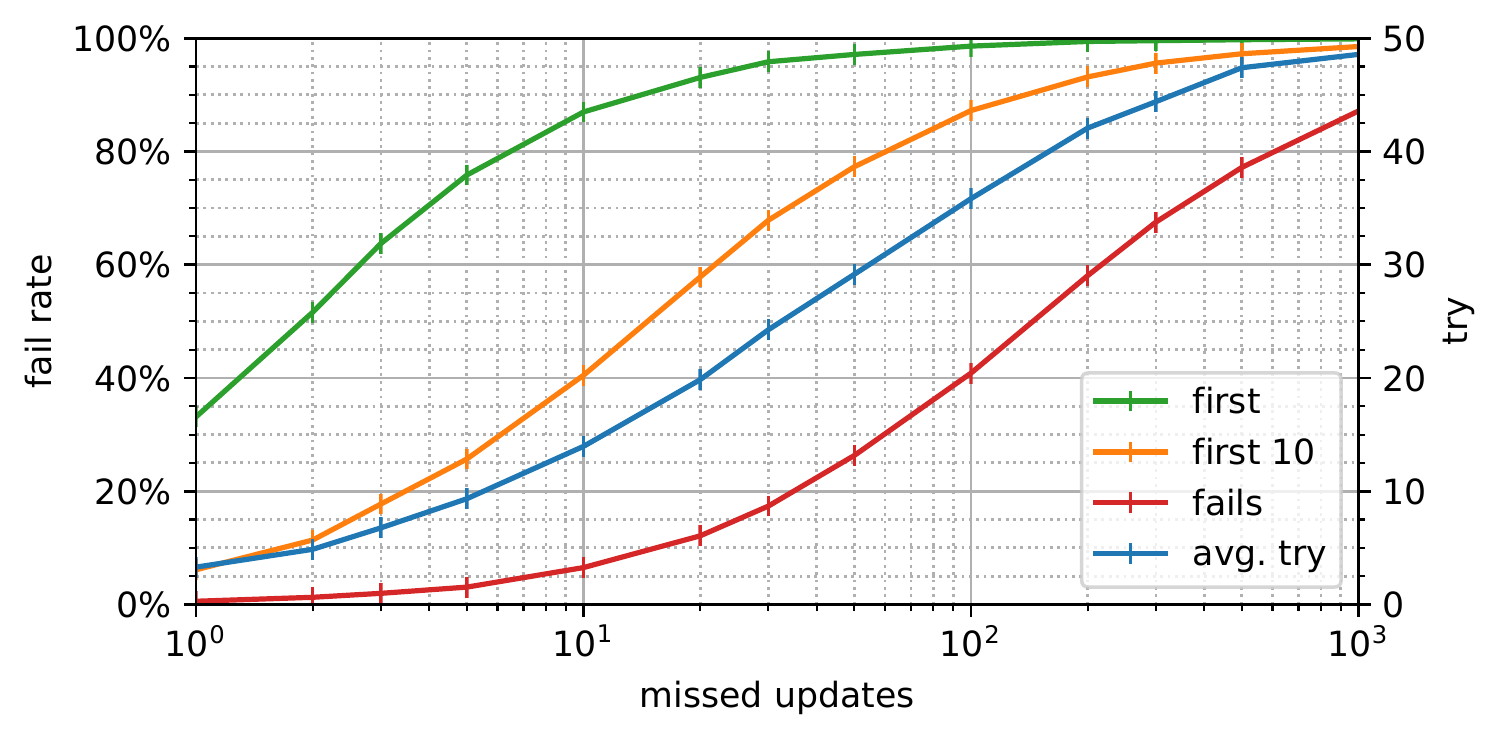}
	\caption{Simulation results for the random \ac{poi} repair for increasingly higher number of missed updates.}
	\label{fig:eval_prob_rndnodes}
\end{figure}

\Cref{fig:eval_prob_rndnodes} shows the simulation results for this strategy with an increasingly higher number of missed updates. 
\emph{first} depicts the percentage of runs which failed to repair the outdated \ac{poi} on the first try and \emph{first 10} for the first ten tries.
After trying 100 \acp{poi}, the run gave up and consider the attempt failed.
Thus, \emph{fails} shows the percentage of runs that stopped after 100 tries.
Finally, \emph{avg. try} sketches the average number of \acp{poi} it took to successfully repair the outdated \ac{poi}.

\begin{figure}[h]
	\centering
	\includegraphics[width=1\columnwidth,trim=0.25cm 0.25cm 0.25cm 0.2cm, clip]{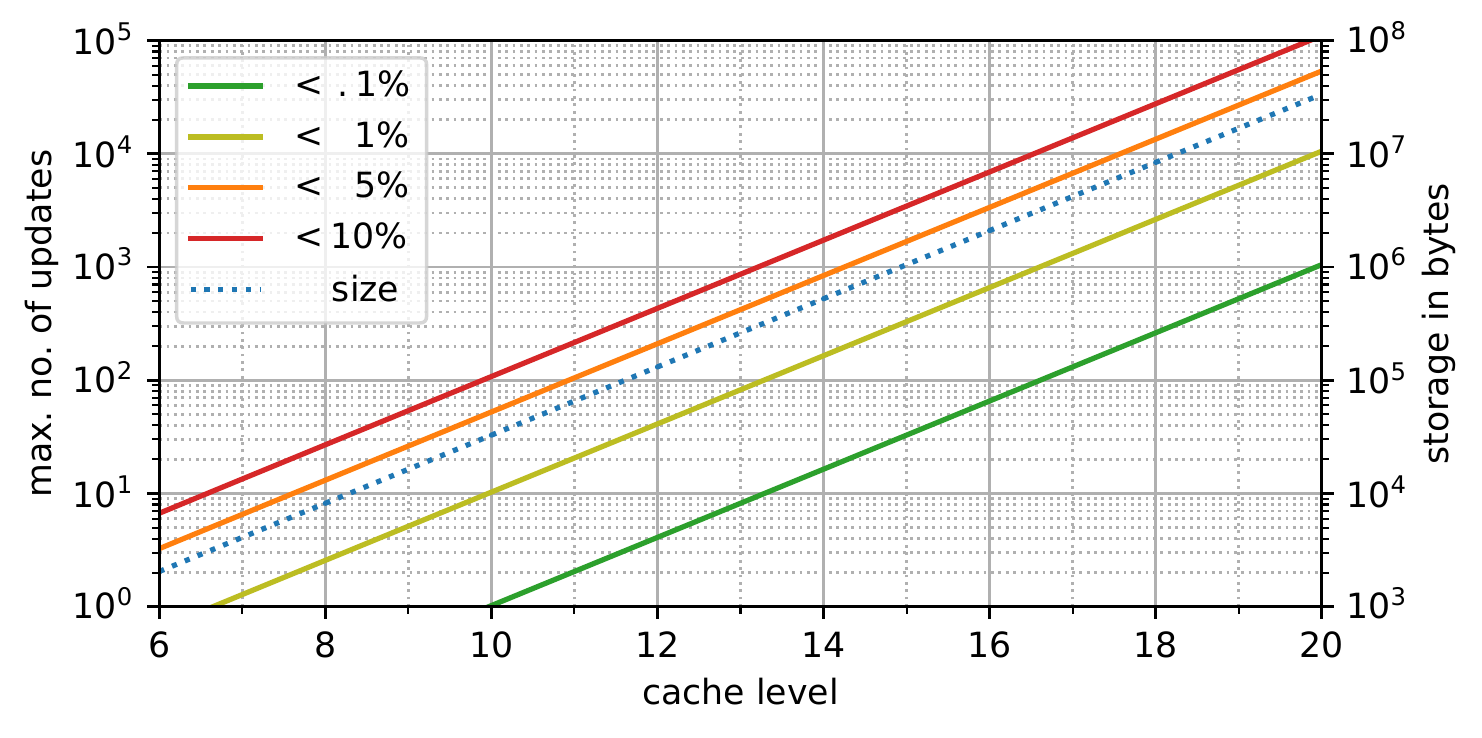}
	\caption{Theoretical number of missable updates for target fail probabilities regarding \emph{LC}-based repair.}
	\label{fig:eval_prob_theory}
\end{figure}

\paragraph{Level-Cache Repair Analysis}
From a theoretical perspective, each \emph{LC} divides each \epochtree into 2\textsuperscript{\emph{clvl}} parts.
An outdated \node should be able to successfully use an up-to-date \emph{LC} to repair its \ac{poi}, if no updates happened in the same part as its certificate is located in.
This model is shown in \Cref{fig:eval_prob_theory}.
We assume that each missed update has a 1/(2\textsuperscript{\emph{clvl}}) chance to be in the same part as an outdated \node, as the distribution of leaf positions is uniform (cf. \Cref{sec:smt}).
The differently colored lines show how many missed updates a \emph{LC} can handle, with a probability less than the target percentage.
The dotted line shows the storage overhead for each cache level.
Our simulations confirmed the theoretical suggestion without significant deviations.

\subsection{Performance Evaluation}
\label{sec:eval_perf}
In this section, we evaluate the performance of \sysname.
First, in terms of run-time for low-power devices and second, the effectiveness for large-scale applications.

\subsubsection{In-Orbit Setup}
\label{sec:eval_runtime}
We implemented a prototype of \sysname to evaluate run-time performance of all basic operations and distributed repair. 
The prototype is written in Python.
While Python is not optimal in terms of performance on constrained devices, we believe that it suffices for a prototype.
Further, we used Python's \texttt{hashlib} for hash computations, which internally calls the OpenSSL native library~\cite{pythonhashlib}.
Hash computations constitute the majority of computational overhead in \sysname.
The prototype is available on GitHub\footnote{\url{https://github.com/vcer4pki/VCER}}.

To show the feasibility for one of the most challenging constrained networks, satellite networks, we deployed the prototype on ESA's OPS-SAT satellite, an in-orbit platform that is open for any party to register and upload experiments~\cite{opssat}.
For this, we adjusted our prototype's code to run as a NanoSat MO Framework app~\cite{nmf}, a straightforward way to run code on OPS-SAT, as it provides crucial services, such as data en-/decoding for transmissions.
For this, we had to convert our prototype to Java, a requirement to use the MO Framework.
The Java prototype uses the internal \texttt{BigInteger} class for the bit-operations and the internal \texttt{MessageDigest} class for the hash calculations.
On the ground, we ran \ca and deployed two additional devices, which were connected to ESA's satellite dish via a SSH tunnel to transmit data from and to the satellite.
OPS-SAT uses the S-Band frequencies as the main link for data communication, with up to 256 kbit/s for communication from the satellite and 1 Mbit/s to the satellite~\cite{opssat}.
Due to OPS-SAT's polar orbit and a single satellite dish in central Europe, a rough estimate for the amount of communication opportunities is less than 6 passes per day, each with less than 10 minutes of varying bandwidth capacity.
This allowed us to successfully test \sysname's feasibility on a deployed satellite with constrained communication, including revocation checks, uploading \ca updates, and distributed repairs between devices.

In the following, we present performance measurements over 1000 executions for the most complex aspects of \sysname.
We use SHA-256 as a hash algorithm and \emph{secp256r1} ECDSA for signatures.
OPS-SAT runs on a Altera Cyclone V SoC with 800MHz and 1GB of RAM, while on the ground one device runs on a Raspberry Pi 3B+ with 1.4 GHz and 4GB of RAM, the other on a Raspberry Pi Zero W with 1GHz and 512MB of RAM.
Note that OPS-SAT also runs other crucial systems in parallel, such as the Attitude Determination and Control System (ADCS).
Furthermore, while OPS-SAT is quite powerful compared to other deployed satellites, upcoming satellite hardware will be more powerful, e.g., see the DAHLIA project~\cite{dahliaproject} or the RAD5500~\cite{baerad}.
Results in \Cref{tab:eval_runtime} indicate that even with a much weaker (in terms of performance and RAM) system the execution of the individual steps is still fast, keeping most execution times in the range of approximately less than 40ms.
An exception is processing 20 update \acp{poi}, which is quite a high number for a single epoch and expected to be rare.
This makes \sysname practical even for low-power devices, and thus fulfills requirement \ref{r:4} for low-performance devices.

\begin{table}
	\caption{Run-time measurements of \sysname, @A refers to the Raspberry Pi 3B+, @B refers to the Rasbperry Pi Zero W, @C refers to the OPS-SAT satellite.}
	\resizebox{\columnwidth}{!}{%
		\def\arraystretch{1.2}
		\begin{tabular}{|l|d{3.3}|d{3.3}|d{3.3}|}
			\hline
			\textbf{Operation} & \multicolumn{1}{c|}{\textbf{@A [ms]}} & \multicolumn{1}{c|}{\textbf{@B [ms]}} & \multicolumn{1}{c|}{\textbf{@C [ms]}}\\\hline
			Signature check \& \primem exchange & 2.915 & 6.869 & 31.008 \\
			\ac{poi} authentication & 3.776 & 19.847 & 228.083 \\
			Processing 20 update \acp{poi} & 76.929 & 407.744 & 4544.915 \\
			Single \ac{poi} repair & 7.574 & 39.377 & 452.260 \\
			\emph{LC} (\emph{clvl} = 7) repair & 6.191 & 29.474 & 264.278 \\
			\hline
		\end{tabular}
	}
	\label{tab:eval_runtime}
\end{table}

\subsubsection{Large-Scale Performance}
\label{sec:eval_sim}
To evaluate \sysname in terms of scalability, we implemented a large-scale network simulation in Python, directly leveraging the prototype described in \Cref{sec:eval_runtime}, running on a Intel Xeon CPU E5-2650 v3 @ 2.30GHz with 10 cores and 256GB of memory.
The simulation runs are executed with an increasing number of \nodes, from $10\,000$ up to a million, over the course of 4 simulated weeks.
In line with related work, we simulate the aspects that affect \sysname's efficiency, i.e., communication overhead over the entire network.
For instance, delays for individual communications between nodes will not significantly affect the system.
As with the accompanying example in \Cref{sec:design}, we use the SHA256 hash function and \emph{secp256r1} ECDSA for signatures.
Further, \forest is split into 52 epochs, representing 52 weeks of a lifetime of one year per certificate, and 7 parities consisting of 2 Bytes for each \primeparity.

Each simulated week, an epoch change occurs, which executes the forest prune (cf. \Cref{sec:forest}) and issues $0.1\%$ new certificates (\textasciitilde 5\% yearly).
On each day in the simulation, $0.028\%$ of the active certificates are revoked (\textasciitilde 10\% yearly), which are re-issued the next day in the newest epoch.
For example, as \acs{iot} devices are notorious for having security bugs, we assume a relatively large share of yearly revocations.
Our share of revocations is based on the estimate of certificates affected by the Heartbleed bug in the Internet~\cite{webpkimeasure}, one of the biggest recorded revocation event.
Splitting the yearly revocations in days is in-line with related work \cite{letsrevoke}, as even in the case of Heartbleed, revocations occurred gradually~\cite{webpkimeasure}.

Every time \ca distributes an update, a random share of \nodes do not receive the update, i.e., become outdated.
We simulated different missing shares of 10\%, 30\%, and 50\% in separate runs.
10\% of \nodes are cachers that additionally store a \emph{LC} with \emph{clvl} = 7; yet, also can miss updates.
For outdated \nodes to repair their own validation information, each \node encounters 5 random \nodes each hour.
During those encounters, both \nodes exchange \primem, update their \forest if applicable as well as notice outdated validation information, and if only one \node is outdated try the distributed repair.
After an outdated \node has met 30 up-to-date \nodes and was not able to repair its validation information, it will give up and request it directly from \ca.

\begin{figure}[h]
	\centering
	\includegraphics[width=0.9\columnwidth,trim=0.25cm 0.25cm 0.25cm 0.2cm, clip]{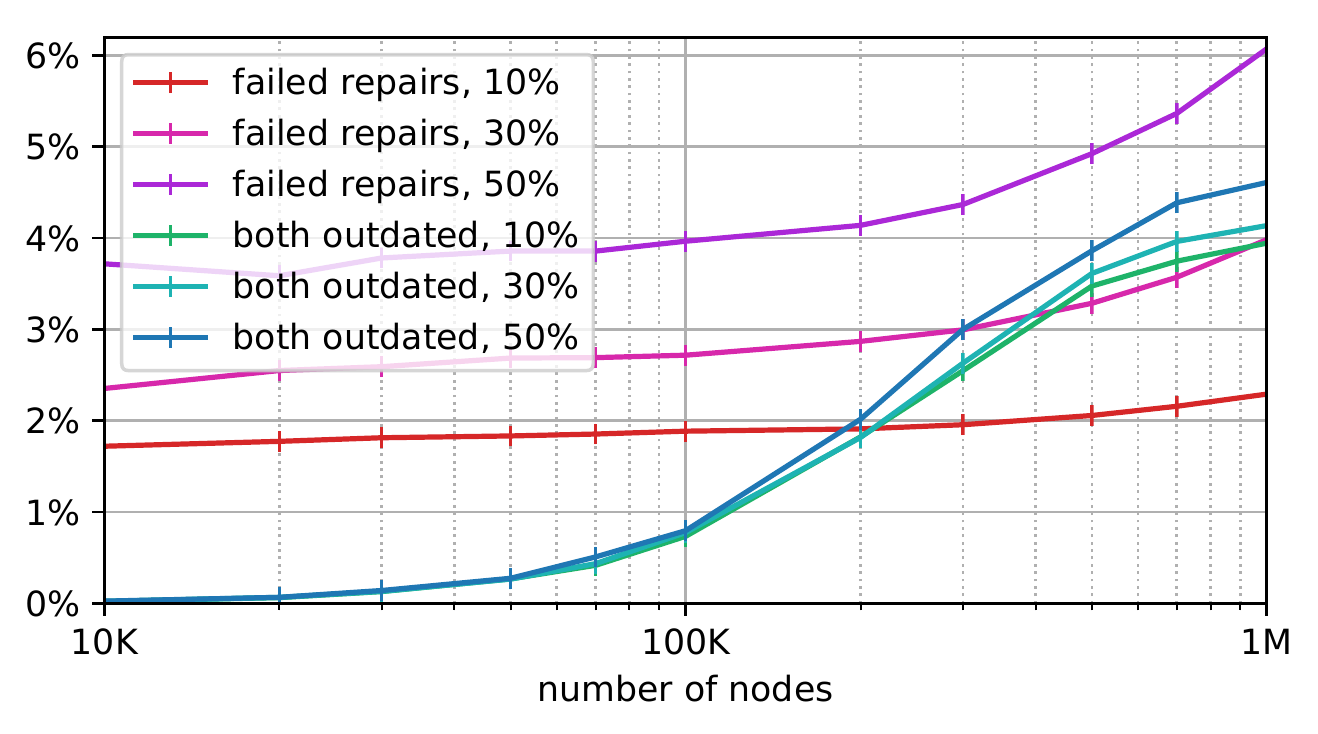}
	\caption{Simulation results for percentage of \nodes failing to repair their \ac{poi}, when they missed \ca updates, and the percentage of all encounters where both \nodes are outdated; both for different missing shares}
	\label{fig:eval_sim_perc}
\end{figure}

\Cref{fig:eval_sim_perc} shows the measurements of two failure percentages.
The \emph{failed repairs} shows the share of outdated \nodes, which were not able to distributively repair their validation information and instead, required a direct request to \ca.
Even in the drastic case of 50\% \nodes missing updates with 1 million \nodes total, more than 93\% of the outdated \nodes are able to collaboratively repair their \ac{poi}.
Therefore, \sysname fulfills requirement \ref{r:1} and \ref{r:5} for a \node's own validation proof.
Outdated \nodes need to meet 8.9, 9.3, and 10.1 up-to-date \nodes on average until the distributed repair is successful, for 10\%, 30\%, and 50\% of \nodes missing updates respectively.
Assuming only one of the two \nodes in an encounter needs an up-to-date \ac{poi}, e.g., to establish a secure channel without mutual authentication, we also measured the percentage of encounters in which both \nodes are outdated.
With 1 million \nodes and 50\% \nodes missing updates, there are less than 5\% of encounters, in which both do not have an up-to-date \ac{poi}.

\begin{figure}[h]
	\centering
	\includegraphics[width=0.9\columnwidth,trim=0.25cm 0.25cm 0.25cm 0.2cm, clip]{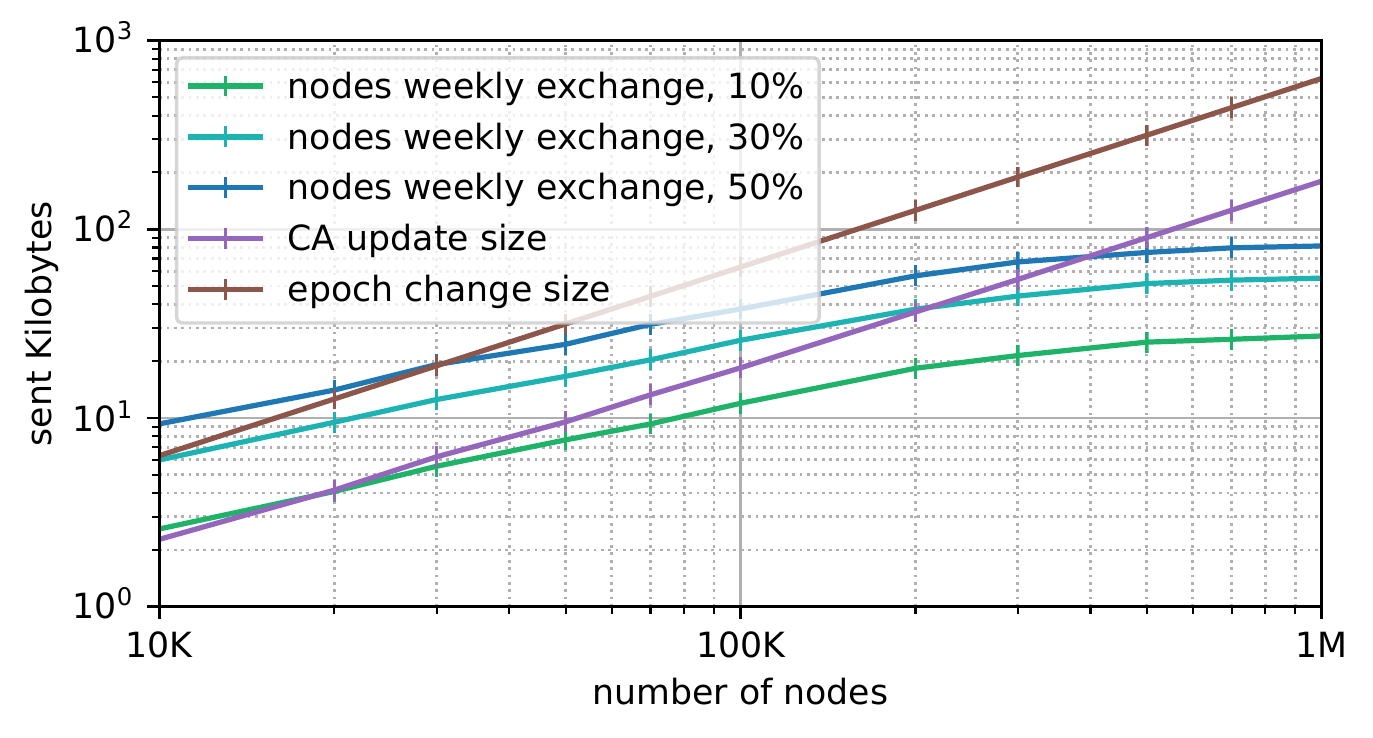}
	\caption{Simulation results for sent out Kilobytes for daily \ca updates, weekly epoch change update size, and average weekly exchanged communication between \nodes for different missing shares.}
	\label{fig:eval_sim_sent}
\end{figure}

\Cref{fig:eval_sim_sent} depicts three key communication overhead measurements.
The graphs for \emph{nodes weekly exchange} represents the average amount of exchanged Kilobytes directly between \nodes, per node per week.
This includes the \primem exchange, \forest repair, and distributed repair of outdated \acp{poi}.
With 1 million \nodes and 10\% of \nodes missing updates, \nodes directly exchange 27.2KB per week and 81.6KB with 50\%.
The \emph{CA update size} is the average update size sent out daily by \ca.
This contains both the revocations as well as the re-issuance the day after revocation (0.056\% of total number of \nodes).
This metric is independent of the missing share and for 1 million \nodes, the daily \ca update is 179.3KB on average.
Finally, the \emph{epoch change size} shows the size of the weekly update sent out by \ca.
Independent of the missing share as well, the epoch change is 629.8KB.
Note, as mentioned in \Cref{sec:ca_updates}, that the epoch change update does not need to be distributed to all of the network and only needs to be sent to \nodes affected by epoch change.

%% file: sections/discussion.tex
\section{Discussion}
\label{sec:discussion}
This section discusses various practical aspects of \sysname.

\paragraph{PKI integration.}
Since \sysname only deals with certificate validation, it needs a \ac{pki} as the basis for deployment.
To the best of our knowledge, \sysname can be used alongside any \ac{pki} scheme, similar to OCSP Stapling, 
by having the prover provide validation evidence (see \Cref{sec:comparison}).
Ideally, the authentication \ac{poi} is directly sent along with the certificate 
during the handshake protocol to reduce communication overhead.
Furthermore, to allow audit of individual changes, \ca can keep a log of all individual operations, 
such as revocation of a certificate. External observers need only to store 
the resulting \primem for each \ca update, allowing them to verify \ca logs after the fact.

\paragraph{Multiple Certificate Authorities.}
While we assume a single \ac{ca} throughout the paper, in some real-world use-cases 
there multiple \acp{ca} can be involved.
This can be accommodated by employing a committee that acts as a single \ca 
via consensus among the members, for all operations.
For example, a signature scheme to allow for such a strategy are BLS signatures~\cite{bls}, which can 
model the necessary threshold-based signatures to create \primesig, verifiable by \nodes.
Alternatively, each member can maintain its own \forest and be responsible for certifying their distinct group of nodes in the network.
In such a case, a node from one vendor can simply collect and keep the most recent \forest of other vendors to authenticate their nodes.
\nodes would only have to deal with \ac{poi} updates regarding their vendor's \forest, while only keeping the other \forest up-to-date.

\paragraph{Caching Certificates.}
If a \node regularly communicates with another \node, the other \node's certificate and 
\ac{poi} can be cached to avoid subsequent redundant checks.
A \node can keep cached \acp{poi} up-to-date, just like it does for its own \ac{poi}.
This way, there is a good chance that cached certificates are ready for use, even after many 
updates. If any cached \acp{poi} get revoked, they can be discarded.
Furthermore, since a \node can provide many up-to-date \acp{poi} to an outdated node, 
this strategy can vastly improve the direct \ac{poi} repair approach, as discussed in \Cref{sec:directrepair}).

\paragraph{Dynamic Cache Sizes.}
While we assumed uniform use of \emph{LC}, i.e., all cacher \nodes use the same \emph{clvl} 
across all epochs in \forest, it may be advantageous in some scenarios to use a dynamic cache size.
On the one hand, cacher \nodes with bigger storage can also keep bigger caches, allowing them 
to succeed more often when executing distributed repairs. On the other hand, the same 
cacher may use different cache sizes for different epochs. For example, if most updates are 
expected within newer epochs, a cacher can use a bigger cache for these than for almost 
expired epochs. Note, a bigger \emph{LC} can easily construct a smaller one, e.g., when 
transitioning to a smaller cache on an epoch change.

%% file: sections/comparison.tex
\section{Comparison to OCSP and CRL schemes}
\label{sec:comparison}
This section focuses on comparing \sysname to the most commonly found types of schemes found in the revocation space, \ac{ocsp}, \acp{crl}, and variants that improve on \acp{crl}.

\paragraph{OCSP Stapling.}
In \ac{ocsp}, the verifier of a certificate directly requests a confirmation from \ca that the certificate is not revoked.
Due to the on-demand nature of \ac{ocsp}, it is not directly applicable to our system model (cf. \Cref{sec:system_model}), as it requires a fast and reliable connection to \ca to properly function.
Nevertheless, there is also \ac{ocsp} Stapling.
Here, the proving party requests the \ac{ocsp} for its own certificate, stores it, and directly delivers it to the verifier.
Each \ac{ocsp} has a validity period that enforces a fresh \ac{ocsp} eventually.
A single \ac{ocsp} request is around 4KB in size~\cite{webpkimeasure}.
However, if we assume the same validity period as in our accompanying example of a single day, this means all \nodes in the network each needs to request a new \ac{ocsp} daily.
For example, even with $100\,000$ \nodes, this would already amount to around 390MB of communication overhead across the network, compared to our 18.4KB via the universal \ca update.

\paragraph{Traditional CRLs.}
With the \ac{crl} approach, \nodes need to maintain and store the entire \ac{crl}.
However, these lists can grow very large.
For example, in the Internet the \acp{crl} size for the median certificate is 51KB, ranging up to 76MB for a single \ac{crl}~\cite{webpkimeasure}.
Yet, even assuming an optimal \ac{crl}, i.e., storing only strictly necessary entries, would already require over 3.6MB of storage to cover for the whole year with 1 million \nodes (as in our evaluation)---assuming an average of 38 Bytes per entry~\cite{webpkimeasure}.
In contrast, \nodes in \sysname only need to store \textasciitilde3KB to achieve the same.

\paragraph{Enhanced CRLs.}
While traditional \acp{crl} can be very large, there are numerous works to significantly reduce the storage and update overhead.
To the best of our knowledge, the most notable works in this space recently are CRLite~\cite{crlite} and Let's Revoke~\cite{letsrevoke}.
Both aim to reduce storage and update sizes of revocation information for the Internet.
CRLite uses a cascade of decreasingly smaller bloom filters to aggregate revocation information.
For 1 million certificates, this requires \textasciitilde112.5KB of storage~\cite{crlite}.
Delta updates model the difference between the old and the new bloom filters on all levels, which can be applied in a XOR-like fashion.
Modeling a change of $0.056\%$ in the contained certificates, e.g., on daily revocations plus re-issuance (as in our evaluation), requires around 10\% to 50\% of the original filter size (according to the measurements presented in Figure 6 in \cite{crlite}).
Let's Revoke~\cite{letsrevoke} uses a bitvector per expiration day per CA, which flags if consecutively numbered certificates have been revoked.
As these bitvectors are expected to have many zeroes, they can further be compressed to save storage space.
With 1 million certificates and 10\% of them revoked, this requires \textasciitilde70KB of storage (\textasciitilde125KB when not compressed)~\cite{letsrevoke}.
For delta updates, bitvectors are constructed that can be applied to the original bitvectors via simple bit operations.
A compressed update modeling $0.056\%$ of certificate changes is \textasciitilde2KB in size~\cite{letsrevoke}.
Note, these schemes are specifically designed for the Web's \ac{pki} and are not concerned with constrained networks.

Both approaches require nodes that missed updates to directly contact any authority or their delegates (e.g., \emph{aggregators} in \cite{crlite}) to keep up-to-date.
In this case, outdated nodes need to individually request the specific range of updates they missed.
On the one hand, this requires careful placement of delegates regarding the targeted topology to ensure coverage, which is difficult in many constrained networks, e.g., due to the nature of a dynamic topology.
On the other hand, all affected nodes would need to individually request missed updates.
For example, a 2KB delta update in Let's Revoke~\cite{letsrevoke} for 1 million nodes with 10\% of nodes missing a \emph{single} update would surmount to 195MB of communication overhead across the network.

In contrast, \sysname only requires less than 3KB of storage on each node and uses universal updates that can be applied at any time, e.g., even though a node missed an update, it may receive a subsequent update in the meantime, which may repair its proof without further requests.
Nodes in \sysname can also help each other to distributively repair their proof, eliminating the need for individual requests to the CA for the principal share of nodes (as shown in \Cref{sec:eval_sim}).
Further, even if the distributed repair fails for nodes, with the up-to-date and quickly disseminated \forest they can still correctly validate other certificates.
Finally, when nodes \emph{do} need to request their fresh \ac{poi} from the CA, it only requires less than 1KB of communication overhead each.

%% file: sections/relatedwork.tex
\section{Related Work}
\label{sec:related_work}
In the following, we will examine the related work in the field of certificate validation.
Aside the ones mentioned throughout this section, there are also works that focus on constructing \acp{pki} for \ac{iot}; yet, for revocation checks, they rely on \acp{crl}~\cite{colliotpki}, on-demand checks~\cite{pki4iot} that we both discussed in \Cref{sec:comparison}, or are blockchain-based~\cite{iotbcpki,iotbcpki2}, which we examine later in this section.
Otherwise, recent works are dominated by observer-based approaches among \acp{ca}, end-user-based approaches, or blockchain-based approaches.
Finally, we compare preceding works focusing on efficient validation for untrusted validation directories with \sysname.

\paragraph{Observer-based approaches.}
These works focus on restricting maliciously acting \acp{ca} by monitoring them for suspicious behavior.
Most prominent in this area is Certificate Transparency (CT)~\cite{ct}, already adopted by many \acp{ca} and browsers for the Internet.
In this approach, an append-only Merkle Tree is used to log all issued certificates by a \ac{ca} as hash leaves.
On issuing, the \ac{ca} publishes the new certificates along with a \emph{consistency proof}, which proves that the previous Merkle Tree is contained in the now extended tree.
Observers, such as other \acp{ca}, check if the new certificates contain any unjustly issued ones, e.g., for domains that the issuing \ac{ca} is not responsible for.
The consistency proof ensures all certificates were published and correctly appended to the \ac{ca}'s Merkle Tree.
On the end-user side, aside from validating the certificate directly with the issuing \ac{ca}, the client additionally requests the \ac{poi} of the certificate from multiple observers.
Falsely issued certificates then become apparent; yet, this does not cover revocations.
An informal report by Laurie and Kasper hints at extending CT with \aclp{smt} to provide \emph{Revocation Transparency}~\cite{smt}.

Enhanced Certificate Transparency~\cite{enhanced-ct} aims to extend the CT approach to also handle revocations efficiently.
The authors argue that search in the Revocation Transparency proposal~\cite{smt} remains linear in the number of issued certificates.
Aside an append-only Merkle Tree as in CT, the paper introduces an additional tree that is ordered by the subject identities, allowing for logarithmic look-up of revocations.
Nevertheless, to ensure consistency, each observer still needs to verify all certificates published and their inclusion into the respective trees. 
The paper mentions its approach can be extended in a distributed manner, so users require less trust into the \acp{ca}.
However, this would require random monitoring of \acp{ca} by all users as well as gossiping the observed information to identify inconsistencies.
Further work focuses on improving resilience of these approaches against colluding \acp{ca}, such as AKI~\cite{aki} or ARPKI~\cite{arpki}.

While these observer-based approaches use similar cryptographic structures to \sysname, i.e., hash trees and \acp{poi}, they aim at the orthogonal goal of limiting malicious behavior of \acp{ca} by giving them the means to efficiently monitor each other.
\sysname, on the other hand, provides efficient certificate validation for end users among each other.
Further, \sysname is not reliant on having reliable connectivity to trusted parties for validation.

\vspace{-0.1em}
\paragraph{End-user-based approaches.}
These approaches aim to modify the observer-based schemes to allow end-users to monitor for inconsistencies regarding their certificates themselves.
In CONIKS~\cite{coniks}, a Prefix Merkle Tree is used, along with randomly generated user IDs, to protect the privacy of users.
Here, all users need to constantly check all issuing parties for certificates of their own domain, to recognize abuse.
When the issuing \ac{ca} updates its tree in any way, it also needs to re-issue all of its users' \acp{poi} so they are correct with respect to the new tree root.
Revocation is handled by on-demand requests to the respective \ac{ca}.
DTKI~\cite{dtki} aims to provide users full data ownership for their certificates by introducing a Merkle Tree log per individual domain.
Each log is maintained in a decentralized database, based on consensus of multiple independent entities.
Users need to gossip all tree roots among each other to prevent problems on network partitioning.
Additionally, DTKI revocation is also done with on-demand requests.

These approaches provide users capabilities to monitor their own domains.
However, they are very communication heavy and rely on on-demand requests to check for revocation, making them unsuitable for constrained networks.
In contrast, \sysname requires minimal communication by only distributing a number of hashes, i.e., \forest.
Further, it allows nodes to collaboratively repair their individual validation proof, mostly without the need to contact any authorities.

\vspace{-0.1em}
\paragraph{Blockchain-based approaches.}
The works in this area aim to shift trust from the \acp{ca} to the blockchain.
One of the first proposals in this area is Certcoin~\cite{certcoin}, which builds a decentralized \ac{pki} based on Namecoin~\cite{namecoin}.
The idea is to simply store all certificate updates, including revocation, on the blockchain.
A certificate owner may send the \ac{poi} for the block containing the certificate for validation, similar to Bitcoin's Simplified Payment Verification~\cite{bitcoinspv}.
This requires a user to monitor and store all block headers, instead of the full blockchain.
Analogously, Blockstack~\cite{blockstack} directly builds on Bitcoin for improved security and separates different abstraction layers for simplified access.
For certificate validation, a user contacts multiple full nodes (i.e., storing the entire blockchain) and checks if the responses are consistent with each other.
EthIKS~\cite{ethiks} simply puts CONIKS~\cite{coniks} on top of Ethereum~\cite{ethereum} and uses smart contracts for global monitoring and validation of certificates.

All these approaches share the notion of having, at least indirect, access to the blockchain.
This is difficult to guarantee in constrained networks or requires significant storage overhead on limited devices.
On the contrary, \sysname only requires minimal storage overhead, without the need to be constantly connected to any specific nodes.

\vspace{-0.1em}
\paragraph{PKI for dynamic networks.}
There are approaches specifically aiming to construct \acp{pki} for constrained networks.
There are many works focusing on Mobile Ad-Hoc Networks~\cite{manetself,manetdistr,manetmoca,manetneighbor} and Delay-Tolerant Networks~\cite{dtncrlrevoke,dtnsocial,dtnibc}.
Note, the latter is considered the state of the art for satellite networks.
They generally aim to distribute the role of the \ac{ca} among the network of nodes.
This usually requires a lot of coordination between all nodes.
However, the key aspect in the context of this work is how revocation is handled.
For this, the approaches either rely on traditional \acp{crl}~\cite{manetdistr,manetmoca}, individual revocation information aggregated by exchanging them among the network~\cite{manetself,manetneighbor,dtncrlrevoke}, or by simply limiting the lifetime of certificates~\cite{manetself,dtnibc}.

To the best of our knowledge, all schemes in these areas either share similar problems of \acp{crl}, require communication heavy coordination between nodes, or introduce an additional vulnerability window.
In \sysname the vulnerability window can be kept to a minimum, as it relies on distributing only \forest.
Otherwise, it minimizes the communication overhead required, making it well suited for constrained networks.

\vspace{-0.1em}
\paragraph{Efficient Validation Directories.}
There have been several works aiming to provide efficient certificate validation~\cite{kocher1998certificate,naor2000certificate,gassko2000efficient,novomodo,quasimodo}.
They use authenticated data structures providing efficient validation proofs, to allow the use of untrusted directories capable of answering validation requests, while removing the need for distributing entire \acp{crl}.
The reduced communication overhead enables shorter validity periods of revocation information, e.g., daily or even hourly.
The data structures are either based on trees for revocation~\cite{kocher1998certificate,naor2000certificate}, trees covering both revocation and validation~\cite{gassko2000efficient}, hash-chains containing all validity periods~\cite{novomodo}, or a combination of both hash-chains and trees~\cite{quasimodo}.

These approaches aim at supplying directories with fresh validation information by regular updates from the \acp{ca}, e.g., by distributing tree updates~\cite{kocher1998certificate,naor2000certificate,gassko2000efficient}.
In contrast, \sysname supplies all nodes with fresh validation information by only distributing \forest.
Nodes can additionally collaborate without any special directories to keep their proofs up-to-date.

%% file: sections/conclusion.tex
\vspace{-0.1em}
\section{Conclusion}
\label{sec:conclusion}
In this work, we presented \sysname, a novel certificate validation scheme, which is designed to work in a distributed and efficient manner.
\sysname can be used to augment any \ac{pki} scheme, enabling it to work even in constrained networks.
This is achieved by introducing data structures and operations that allow for fast dissemination of validation information as well as a collaborative way for nodes to keep up-to-date.
We have demonstrated the efficacy and efficiency of \sysname with large-scale simulations modeling a constrained network.

\section*{Acknowledgments}
The authors of the Technical University of Darmstadt were supported by the European Space Operations Centre with the Networking/Partnering Initiative and the European Union’s Horizon 2020 Research
and Innovation program under Grant Agreement No. 952697 (ASSURED). 
Gene Tsudik's work was supported in part by NSF awards SATC-1956393 and CICI-1840197, as well as a subcontract from Peraton Labs.

%% file: sections/appendix.tex
\appendix
\section{In-depth Operation Description}
\label{sec:indepth_algos}
This section shows the in-depth working of the basic operations for epoch trees.

\subsection{Preliminaries and Notation}
\label{sec:notation}

\Cref{tab:notation} summarizes our notation for the operations.
When addressing a position in the \ac{smt} the term \emph{depth} or \emph{depth level} is used, where a higher number depth means closer to the leaves and a lower number depth means closer to the root.
This is also illustrated in \Cref{fig:smt} (b).
When describing a for-loop in the algorithms, note that ``\textbf{for} $i \gets 0$ \textbf{to} $x$ \textbf{do}'' means $i$ gets assigned $[0,x)$ in the course of the loop.
For brevity, we skip the checks for empty hashes in \emph{EmptyHashesList} and simply check for $\varnothing$.

\begin{table} [h]
	\caption{Variables and operation definitions.}
	\normalsize
	\def\arraystretch{1.2}
	\begin{tabularx}{\columnwidth}{ll} 
		\hline
		\multicolumn{2}{l}{\textbf{Variables}} \\ 
		\hline
		$LUT$ & Look-Up-Table storing hashes of all \\
		& non-empty branches and leaves of $SMT$ \\
		\textbar\texttt{H}\textbar & Bit length of a hash digest and depth of $SMT$ \\
		\textbar$l$\textbar & Number of elements in list $l$ \\
		\emph{LC} & Level-Cache-List sorted by position of \\
		& the cache elements in $SMT$ \\
		\emph{clvl} & Depth of \emph{LC} \\
		\hline
		\multicolumn{2}{l}{\textbf{Operations}} \\ 
		\hline
		\emph{int}$\langle p \rangle$ & Access bit at position $p$ of \emph{int} from the right \\
		$l$[$p$] & Access element at position $p$ in list $l$ \\
		$LUT$[$p$, $d$] & Get hash at position $p$ at depth $d$, \\
		& the last $d$ bits of $p$ will be ignored, e.g., \\
		& $d=0$ returns $SMT_r$, $d=$ \textbar\texttt{H}\textbar\ returns a leaf \\
		$\sim$$x$ & Flip bit(s) of $x$ \\
	\end{tabularx}
	\label{tab:notation}
\end{table}

\subsection{CA Operations}
\label{sec:basic_smt}

This section describes how \sysname does basic operations.
We use an incremental approach using the Look-Up-Table $LUT$, instead of constructing the tree in one go for a set of leaves~\cite{smt}.
This avoids having to reconstruct the entire epoch tree on an update to it.
As \sysname works by regularly updating the respective epoch trees, this helps to significantly reduce the calculation overhead on the \ca.
This comes at the cost of having to store the $LUT$.

\subsubsection{Add Leaf to \ac{smt}}
\label{sec:add_leaf}
This operation is executed by the \ca and adds one leaf hash to an epoch tree while updating the $LUT$ for subsequent operations.
\Cref{fig:smt} (b) illustrates how it works for $c_1$.
Starting from the leaf, the operation will go up the tree, look-up the respective neighbor in the $LUT$ (see green nodes), calculate and set the intermediate nodes along the leaf's path in the $LUT$ (see green line), and repeat this process until reaching the root.
For the initial epoch tree construction or processing multiple updates, this operation is called multiple times individually for each leaf.

\begin{algorithm}[ht]
	\caption{\texttt{add\_leaf} function for adding a new leaf hash and recalculating the tree while updating the LUT}\label{alg:add_leaf}
	\small
	\raggedright
	\begin{algorithmic}[1]
		\Require \emph{leaf\_hash}, $LUT$
		\Ensure \epochroot, $LUT$
		
		\State $LUT$[\emph{leaf\_hash}, \textbar\texttt{H}\textbar] $\gets$ \emph{leaf\_hash}
		\label{alg:add_leaf:leaf}
		\For{$i \gets 0$ \textbf{to} \textbar\texttt{H}\textbar}
		\State \emph{neighbor} $\gets$ \emph{leaf\_hash}
		\State \emph{neighbor}$\langle$\textbar\texttt{H}\textbar\ $-$ $i \rangle$ $\gets$ $\sim$\emph{neighbor}$\langle$\textbar\texttt{H}\textbar\ $-$ $i \rangle$
		\label{alg:add_leaf:neighbor}
		\If{\emph{leaf\_hash}$\langle$\textbar\texttt{H}\textbar\ $-$ $i \rangle$ = True}
		\label{alg:add_leaf:lookup1}
		\State \emph{left\_hash} $\gets$ $LUT$[\emph{neighbor}, (\textbar\texttt{H}\textbar\ $-$ $i$)]
		\State \emph{right\_hash} $\gets$ $LUT$[\emph{leaf\_hash}, (\textbar\texttt{H}\textbar\ $-$ $i$)]
		\Else
		\State \emph{left\_hash} $\gets$ $LUT$[\emph{leaf\_hash}, (\textbar\texttt{H}\textbar\ $-$ $i$)]
		\State \emph{right\_hash} $\gets$ $LUT$[\emph{neighbor}, (\textbar\texttt{H}\textbar\ $-$ $i$)]
		\EndIf
		\label{alg:add_leaf:lookup2}
		\State $LUT$[\emph{leaf\_hash}, (\textbar\texttt{H}\textbar\ $-$ $i$ $-$ 1)] $\gets$ \texttt{H(}\emph{left\_hash} $||$ \emph{right\_hash}\texttt{)}%
		\label{alg:add_leaf:add}
		\EndFor
		
		\State \textbf{return} $LUT$[$\varnothing$, 0], $LUT$
	\end{algorithmic}
\end{algorithm}

\Cref{alg:add_leaf} describes the operation in detail.
First, it sets the hash to be added on the respective leaf position in line~\ref{alg:add_leaf:leaf}.
It traverses the tree from bottom to top, i.e., starting at the right-most bit in the leaf going successively left.
In line~\ref{alg:add_leaf:neighbor} we construct the position of the neighbor at the current depth by flipping the bit representing this depth.
Depending on which side the leaf and neighbor is, in lines~\ref{alg:add_leaf:lookup1} to \ref{alg:add_leaf:lookup2}, we look-up the left and right hash to concatenate them in correct order, hash them, and set the result at the respective position in the $LUT$ (line~\ref{alg:add_leaf:add}).
This process will be repeated for all depths until the root is reached.
For revocation, the same operation is used regarding the leaf to be removed, except in line~\ref{alg:add_leaf:leaf} we set the leaf to \texttt{H(}$\varnothing$\texttt{)} instead.

\subsubsection{\ac{poi} Construction}
This operation is executed by the \ca to construct a \ac{poi} for the given leaf hash.
The $LUT$, populated by the \texttt{add\_leaf} operation, is used for the \ac{poi} construction.
Only necessary hashes will be put in the \ac{poi} by excluding empty hashes.
However, this additionally requires a \textbar\texttt{H}\textbar-bit sized path bitmap for verification, to know which depth each element in the \ac{poi} represents.

\begin{algorithm}[ht]
	\caption{\texttt{calc\_poi} function for calculating the proof of inclusion for a leaf}\label{alg:calc_poi}
	\small
	\raggedright
	\begin{algorithmic}[1]
		\Require \emph{leaf\_hash}, $LUT$
		\Ensure \emph{path}, \emph{path\_bitmap}
		
		\State \emph{path}[] $\gets \varnothing$
		\State \emph{path\_bitmap} $\gets$ 0
		\For{$i \gets 0$ \textbf{to} \textbar\texttt{H}\textbar}
		\State \emph{neighbor} $\gets$ \emph{leaf\_hash}
		\State \emph{neighbor}$\langle$\textbar\texttt{H}\textbar\ $-$ $i \rangle$ $\gets$ $\sim$\emph{neighbor}$\langle$\textbar\texttt{H}\textbar\ $-$ $i \rangle$
		\State \emph{neighbor\_hash} $\gets$ $LUT$[\emph{neighbor}, (\textbar\texttt{H}\textbar\ $-$ $i$)]
		\label{alg:calc_poi:lut}
		\If{\emph{neighbor\_hash} $\neq$ $\varnothing$}
		\State \emph{path\_bitmap}$\langle i \rangle$ $\gets$ True
		\State \emph{path}\texttt{.append(}\emph{neighbor\_hash}\texttt{)}
		\EndIf
		\EndFor
		
		\State \textbf{return} \emph{path}, \emph{path\_bitmap}
	\end{algorithmic}
\end{algorithm}

The operation is shown in \Cref{alg:calc_poi}.
It works similar to \texttt{add\_leaf}, by starting from the bottom of the tree and working its way up.
For each depth, the neighbor position for the given leaf is calculated and looked-up in line~\ref{alg:calc_poi:lut}.
If the looked-up hash is not empty, it will be appended to the \ac{poi} and the bit at the current depth is set in the path bitmap.
For a revoked leaf the operation does not require any changes.

\subsubsection{\ac{poi} Verification}
For this operation, we calculate the root resulting from a \ac{poi}.
Given the leaf to be verified, its \ac{poi}, and respective path bitmap, the operation can be executed by anyone.
If this root matches the expected epoch root, the certificate represented by the leaf is valid.
Per default, the operation will calculate the root, yet, an optional parameter can be passed to stop at the specified depth-level.

\begin{algorithm}[ht]
	\caption{\texttt{calc\_path\_root} function for calculating the root of a proof of inclusion, optionally only until a given depth}\label{alg:calc_path_root}
	\small
	\raggedright
	\begin{algorithmic}[1]
		\Require \emph{leaf\_hash}, \emph{path}, \emph{path\_bitmap}, [\emph{lvl} $\gets$ 0]
		\Ensure \emph{root\_hash}
		
		\State \emph{result} $\gets$ \emph{leaf\_hash}
		\For{$i\gets 0$ \textbf{to} (\textbar\texttt{H}\textbar\ $-$ \emph{lvl})}
		\If{\emph{path\_bitmap}$\langle i \rangle$ $=$ True}
		\label{alg:calc_path_root:bitmap1}
		\State \emph{neighbor} $\gets$ \emph{path}[0]
		\State \emph{path}\texttt{.popfront()}
		\Else
		\State \emph{neighbor} $\gets$ $\varnothing$
		\EndIf
		\label{alg:calc_path_root:bitmap2}
		\If{\emph{leaf\_hash}$\langle i \rangle$ = True}
		\State \emph{result} $\gets$ \texttt{H(}\emph{neighbor} $||$ \emph{result}\texttt{)}
		\Else
		\State \emph{result} $\gets$ \texttt{H(}\emph{result} $||$ \emph{neighbor}\texttt{)}
		\EndIf
		\EndFor
		
		\State \textbf{return} \emph{result}
	\end{algorithmic}
\end{algorithm}

\Cref{alg:calc_path_root} gives a detailed description of the operation.
The algorithm starts by setting the given leaf in the \emph{result} variable and then works its way up the tree from the bottom.
In the lines~\ref{alg:calc_path_root:bitmap1} to \ref{alg:calc_path_root:bitmap2} it checks if there is an element in the given \ac{poi} for the current depth and if so, it extracts this \ac{poi} element.
Otherwise, an empty hash is assumed instead.
Then both hashes are hashed together in the respective order and set to the \emph{result} variable.
This process is repeated for all depths until we reach the root, or the given depth level, if specified.

\subsubsection{Level-Cache Construction}
\label{sec:lc_construction}
The construction of a \emph{LC} by the \ca is shown in \Cref{alg:construct_lvl_cache}.
Note that the elements of \emph{LC} are sorted left-to-right by their position in the epoch tree.
Thus, we can simply go through all positions in the \emph{LC}, bit-shift it to the very left regarding the digest size, and use this to simply look-up the value at the depth \emph{clvl}.
The \ca can repeat this for all epochs and then distribute the resulting \emph{LC}.
If some \nodes missed this distribution, \nodes that received the \emph{LC} may share it with others, or construct \emph{LC}s with a smaller \emph{clvl} if requested.

\begin{algorithm}[ht]
	\caption{\texttt{construct\_lvl\_cache} for constructing a level-cache with a specified depth}\label{alg:construct_lvl_cache}
	\small
	\raggedright
	\begin{algorithmic}[1]
		\Require \emph{clvl}, $LUT$
		\Ensure \emph{LC}
		
		\State \emph{LC}[] $\gets \varnothing$ 
		\For{$i\gets 0$ \textbf{to} $2$\textsuperscript{\emph{clvl}}}
		\State \emph{position} $\gets$ $i$ $\ll$ $($\textbar\texttt{H}\textbar\ $-$ \emph{clvl}$)$
		\State \emph{LC}\texttt{.append(}$LUT$[\emph{position}, \emph{clvl}]\texttt{)}
		\EndFor
		
		\State \textbf{return} \emph{LC}
	\end{algorithmic}
\end{algorithm}